\documentclass[11pt]{article}

\usepackage[a4paper,margin=2.4cm]{geometry}
\usepackage{amsmath,amssymb,mathtools}
\usepackage{bm}
\usepackage{graphicx}
\usepackage{xcolor}
\usepackage{float}
\usepackage{placeins}
\usepackage[affil-it]{authblk}
\usepackage{hyperref}
\hypersetup{colorlinks=true,linkcolor=blue!60!black,citecolor=blue!60!black,urlcolor=blue!60!black}

\newcommand{\be}{\begin{equation}}
\newcommand{\ee}{\end{equation} }
\newcommand{\ba}{\begin{array}}
\newcommand{\ea}{\end{array}}
\newcommand{\dis}{\displaystyle}
\newcommand{\Det}{{{\rm{Det\,}}}}
\newcommand{\Tr}{{\scalebox{0.9}{${\mathrm{Tr}}$}}}
\newcommand{\kB}{{k_{{\rm{B}}}}}
\newcommand{\cO}{{\cal O}}
\newcommand{\ad}{\mathbf{ad}}
\newcommand{\na}{\nabla}
\newcommand{\rmd}{{\rm d}}
\newcommand{\rd}{{\rmd}}
\def\bX{\mathbf{X}}
\def\bP{\mathbf{P}}
\def\hbX{\mathbf{\hat{X}}}
\def\hbP{\mathbf{\hat{P}}}
\newcommand{\vx}{{\vec{x}}}
\newcommand{\vp}{{\vec{p}}}
\newcommand{\vF}{{\vec{F}}}
\newcommand{\hp}{\hat{p}}
\newcommand{\hx}{\hat{x}}
\newcommand{\hA}{\hat{A}}
\newcommand{\hB}{\hat{B}}
\newcommand{\hH}{{\hat{H}}}
\def\betavar{{\beta_{\scalebox{0.5}{$\varphi$}}}}
\def\omegavar{{\omega_{\scalebox{0.5}{$\varphi$}}}}
\newcommand{\red}[1]{{\color{red} #1\color{black}}}
\newcommand{\blue}[1]{{\color{blue} #1\color{black}}}
\graphicspath{{figures/}}

\title{\bf Attractive statistical forces and Pauli crystal formation in trapped Fermi gases}

\author[1]{Kawon Lee}
\author[1]{Sangeun Oh}
\author[1]{Young Woo Choi\thanks{ywchoi02@sogang.ac.kr}}
\author[1]{Jeong-Hyuck Park\thanks{park@sogang.ac.kr}}
\affil[1]{Department of Physics, Sogang University, Seoul 04107, Korea}

\date{}

\begin{document}
\maketitle

\begin{abstract}
\setlength{\emergencystretch}{1em}
\noindent
Exchange statistics endows identical particles with an effective ``statistical potential'', whose familiar exact form is two-body and purely repulsive for fermions. Here we construct an exact collective many-body form: the thermodynamics of $N$ trapped ideal fermions maps onto classical distinguishable particles governed by a single potential---exactly for harmonic confinement at all temperatures, and to leading semiclassical order for arbitrary potentials. The associated force separates canonically into pairwise contributions; attraction appears already at $N=3$, where its sign obeys an exact geometric criterion. Classical minimization reproduces observed few-body Pauli-crystal symmetries and agrees with the $N=55$ ground-state probability maximum at sub-percent shell accuracy. Heating drives discrete structural transitions of the most probable configuration, accompanied by an ensemble-verified crossover of the strongest force from attractive to repulsive. Both the potential and its forces are directly computable from existing single-shot imaging data, turning quantum exchange into measurable classical mechanics.
\end{abstract}

\vspace{4pt}

\section*{Introduction}

In 1932, Uhlenbeck and Gropper showed that the quantum statistics of an ideal gas can be mimicked classically: two identical particles behave, to lowest order, like distinguishable classical particles interacting through a temperature-dependent \textit{statistical potential}---repulsive for fermions, attractive for bosons~\cite{Uhlenbeck1932}. The concept underlies the exchange corrections to the equation of state and has become a textbook expression of how quantum symmetry masquerades as an effective interaction~\cite{Huang1987,PathriaBeale,MullinBlaylock}. As Huang emphasized, the statistical potential ``originates solely from the symmetry properties of the wave function'' and depends on temperature, so it ``cannot be regarded as a true interparticle potential''~\cite{Huang1987}: it is not a force law of nature but the exact classical bookkeeping of quantum symmetry. Its familiar exact closed form, however, is the two-body result.

Meanwhile, the $N$-body question has become experimental. Deterministic preparation now provides controlled few-fermion systems~\cite{Serwane2011,Murmann2015}; lattice microscopes resolve individual fermions~\cite{Cheuk2015}, and continuum microscopes and matter-wave magnification image particle-resolved configurations~\cite{Brandstetter2025,deJongh2025,Yao2025}. Repeated shots thereby sample spatial or momentum-space correlations up to the $N$-body distribution. The sharpest phenomenon accessed this way is the Pauli crystal~\cite{GajdaEPL2016,Rakshit2017}: geometric order arising purely from antisymmetrization, without any interparticle interaction. Pair-additive applications have summed the two-body statistical potential over all pairs to approximate these structures in two-dimensional traps~\cite{Batle2017,CiftjaBatle2019}. Such models recover selected few-body geometries but retain quantum statistics only as a first-order, pairwise correction; at $N=15$ the resulting shell structure disagrees with the quantum Pauli crystal~\cite{CiftjaBatle2019}. Unlike Wigner crystals~\cite{Wigner1934}, which arise from Coulomb repulsion, Pauli crystals leave the one-body density smooth and featureless; their order lives entirely in higher-order correlations. First observed in 2021~\cite{Holten2021} and recently predicted to trigger zero-threshold cavity superradiance~\cite{OrtunoGonzalez2026}, they are commonly described kinematically, as consequences of antisymmetrization. Their thermal robustness and driven melting have been studied numerically~\cite{Rakshit2017,Xiang2023}. What has remained unavailable is an exact collective $N$-body potential retaining all exchange cycles.

In this work we construct the exact $N$-body statistical potential and develop its consequences. We show that (i)~the thermodynamics of $N$ identical free fermions is \textit{exactly} that of classical distinguishable particles under a collective, configuration-dependent potential $V_{\rm stat}$, with strictly positive Boltzmann weights---the fermionic sign structure is completely absorbed into a determinant; (ii)~in a harmonic trap the mapping remains exact at \textit{all} temperatures with renormalized parameters, and for arbitrary potentials it holds at leading semiclassical order; (iii)~the statistical force admits a canonical pairwise decomposition, closed-form for any $N$, which is purely repulsive only for $N=2$: for $N\geq 3$ attractive pair contributions appear, governed by an exact geometric criterion; (iv)~classical minimization reproduces observed few-body Pauli-crystal symmetries and agrees with the $N=55$ ground-state probability maximum at sub-percent shell accuracy, while its temperature dependence yields discrete structural transitions---gas-like, intermediate, crystalline---together with a crossover of the strongest pair force from repulsive to attractive; (v)~every one of these force-level statements can be evaluated, shot by shot and in closed form, on existing single-shot imaging data.

We emphasize the scope of these statements at the outset. No new interparticle interaction is claimed: $V_{\rm stat}$ is a dual description of free fermions, and its pairwise forces are canonical objects \textit{of that dual}. Their empirical content is that of the configurational statistics they generate exactly---statistics that experiments already measure. Within this honest scope the dual is far from an empty rewriting: it is exact, closed-form, and positive-weight, and it converts questions about antisymmetrized many-body wavefunctions into classical mechanics---minimization, forces, and bonds---that can be computed directly from data.

\section*{Results}

\subsection*{An exact classical dual of the ideal Fermi gas}

For $N$ identical free fermions in $D$ dimensions, with $\beta=(\kB T)^{-1}$, we define the many-body statistical potential
\be
V_{\rm{stat}}(\beta,\bX)=-\beta^{-1}\ln\Det\Big[e^{-\frac{m}{2\beta\hbar^2}(\vx_a-\vx_b)^2}\Big]_{N\times N}\,,
\label{Vstat}
\ee
where $\bX=(\vx_1,\ldots,\vx_N)$ collects the particle positions and the determinant is over the particle indices $a,b=1,\ldots,N$. It satisfies the exact identity (Methods)
\be
\Tr\left(e^{-\frac{\beta}{2m}\hbP\cdot\hbP}\right)=
\frac{1}{N!}\left(\frac{m}{2\pi\beta \hbar^2}\right)^{\frac{DN}{2}}\!
{\dis{\int} D \bX}\, e^{-\beta V_{\rm{stat}}(\beta,\bX)}\,,
\label{ebV}
\ee
with ${\bP\cdot\bP}=\sum_{a=1}^{N}\vp_a\cdot\vp_a$. The right-hand side is a classical configurational integral for \textit{distinguishable} particles interacting through $V_{\rm stat}$, with the Gibbs factor $1/N!$ accounting for particle identity. For $N=2$, Eq.~(\ref{Vstat}) reduces to the Uhlenbeck--Gropper potential~\cite{Uhlenbeck1932,Huang1987}; for general $N$ it is collective: it does not decompose into a sum of pair potentials.

Two structural properties make this dual a legitimate classical ensemble rather than a formal rewriting. First, the matrix in Eq.~(\ref{Vstat}) is the Gram matrix of a Gaussian kernel and is therefore positive definite for any configuration of distinct points: its determinant is strictly positive, $V_{\rm stat}$ is real, and the classical weight $e^{-\beta V_{\rm stat}}$ is positive. The antisymmetry signs of the fermionic wavefunction are thus completely absorbed into $\ln\Det$---in the configurational sector of the trapped ideal Fermi gas there is no sign problem, and the exact quantum configurational statistics can be sampled by standard classical Monte Carlo. Second, as shown below, the dual is not restricted to free fermions: it survives harmonic confinement exactly, at every temperature.

\subsection*{Canonical pairwise decomposition of the statistical force}

The statistical force acting on the $a$-th fermion is
\be
\vF_{a}=-\na_{a}V_{\rm{stat}}={\dis{\sum_{b\neq a}~}}\vF_{b\rightarrow a}\,.
\label{Fa}
\ee
Introducing the symmetric matrix
\be
K_{ab} =\exp\!\left[{-\frac{m}{2\beta\hbar^2}(\vx_a-\vx_b)^2}\right]\,,
\label{Kab}
\ee
the pairwise contribution reads (no sum over repeated indices)
\be
\vF_{b\rightarrow a}=-\vF_{a\rightarrow b}=\frac{2m}{(\beta\hbar)^2}(\vx_{b}-\vx_{a})\, K^{-1\,ab}K_{ab}\,.
\label{PairForce}
\ee
Each contribution acts along the line joining particles $a$ and $b$ and obeys Newton's third law, with the sign of $K^{-1\,ab}K_{ab}$ determining whether it is attractive or repulsive. This decomposition is \textit{canonical}: it is inherited from the determinant representation~(\ref{Vstat}), in which each pair separation enters through a single symmetric matrix entry, and it is the unique central decomposition whose coefficients are the derivatives of that representation with respect to the squared separations (Supplementary Note~7 makes this precise: the decomposition is strictly unique for $N=3$, unique at the representation level for $N\geq 4$, and attraction cannot be eliminated by any continuous, cluster-separable re-decomposition scheme). The pairwise force admits an explicit representation in terms of permutations,
\be
\!\vF_{b\rightarrow a}\!=\!\scalebox{1.2}{$\frac{2m (\vx_{b}-\vx_{a})}{(\beta\hbar)^2e^{-\beta V_{\rm{stat}}(\bX)}}$}\!\left(\!{\dis{\sum_{\sigma}}} (-1)^{\sigma}\delta_{a}^{\sigma(b)}
e^{-\frac{m}{2\beta\hbar^2}\left|\left|{\bX-\bX_{\sigma}}\right|\right|^2}\!\right)\!,
\label{Fba}
\ee
where $\bX_\sigma$ denotes the permuted configuration and $(-1)^\sigma=\pm1$; this form makes the collective, many-body origin of every pair force manifest.

\begin{figure}[!htb]
	\centering
	\includegraphics[width=0.92\columnwidth]{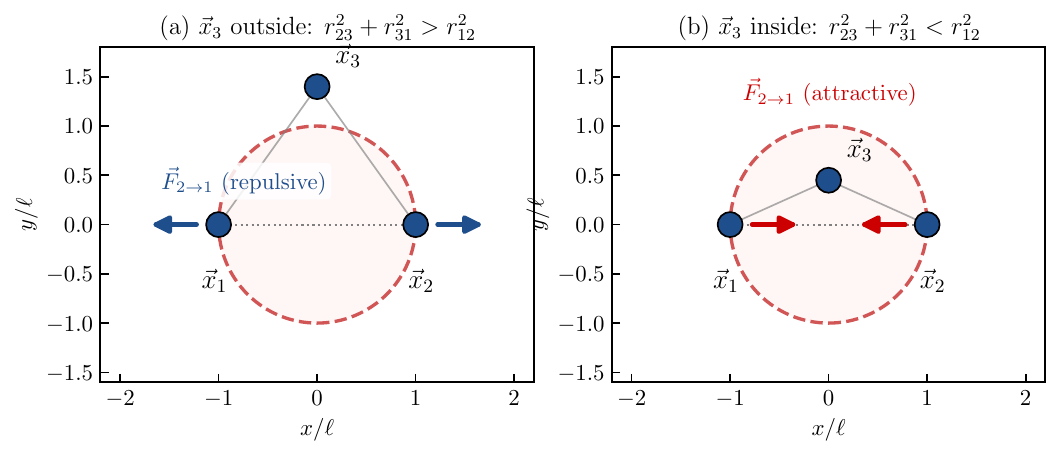}
	\caption{\textbf{Geometric criterion for attractive statistical forces at $N=3$}, shown as a $D=2$ illustration of the dimension-independent condition~(\ref{antipodal}). The force on particle~1 from particle~2 changes sign depending on the position of the third particle relative to the antipodal ball of $\vx_1,\vx_2$ (dashed: the disk with $\vx_1,\vx_2$ as diameter endpoints).
(a)~When $\vx_3$ lies outside the ball ($r_{23}^2+r_{31}^2>r_{12}^2$), the bracket in Eq.~(\ref{N3Force}) is positive and $\vF_{2\to 1}$ is repulsive, reducing to the two-body result~(\ref{N2Force}) as $r_{13},r_{23}\to\infty$.
(b)~When $\vx_3$ lies inside the ball ($r_{23}^2+r_{31}^2<r_{12}^2$), the bracket reverses sign and $\vF_{2\to 1}$ becomes attractive.}
	\label{FIG_N3sch}
\end{figure}

For ${N=2}$, the force reduces to the purely repulsive textbook form~\cite{Uhlenbeck1932,Huang1987}:
\be
\vF^{N=2}_{2\rightarrow 1}=\frac{2m(\vx_{1}-\vx_2)}{(\beta\hbar)^2\left[e^{\frac{m}{\beta\hbar^2}(\vx_{1}-\vx_{2})^2}-1\right]}\,.
\label{N2Force}
\ee
For $N\geq 3$, the force becomes configuration-dependent and can change sign. For ${N=3}$, with $r_{ab}=|\vx_a-\vx_b|$,
\be
\vF^{N=3}_{2\rightarrow 1}=\frac{2m(\vx_{1}-\vx_2) e^{-\frac{m}{\beta\hbar^2}r_{12}^2}}{(\beta\hbar)^2e^{-\beta V_{\rm{stat}}}}\left[1-e^{-\frac{m}{2\beta\hbar^2}\left(r_{23}^2+r_{31}^2-r_{12}^2\right)}\right]\,.
\label{N3Force}
\ee
If the third particle moves far away, $r_{13},r_{23}\rightarrow\infty$, the two-body repulsion~(\ref{N2Force}) is recovered. But when the third particle lies inside the closed ball having $\vx_1$ and $\vx_2$ as antipodal points,
\be
r_{23}^2 + r_{31}^2 < r_{12}^2\,,
\label{antipodal}
\ee
the bracket in Eq.~(\ref{N3Force}) reverses sign and the force between particles 1 and 2 becomes \textit{attractive} (Fig.~\ref{FIG_N3sch}). The criterion~(\ref{antipodal}) is exact, temperature-independent, and holds in any dimension. Statistical forces in many-fermion systems are therefore intrinsically collective and not reducible to pairwise repulsion---the familiar lore ``identical fermions effectively repel'' is the $N=2$ corner of a richer sign structure. For small $N$ the \textit{globally strongest} pair force nevertheless remains repulsive; as shown below, this dominance reverses at sufficiently low temperature for certain particle numbers.

\subsection*{Exact mapping for harmonic confinement and semiclassical generality}

For a harmonic trap, the single-particle thermal kernel is given exactly by the Mehler formula~\cite{FeynmanHibbs,Kleinert} (Methods),
\begin{equation}
\ba{l}
\left\langle x\left|\exp\left[-\beta\left(\frac{\hp^2}{2m}+\frac{m\omega^2\hx^2}{2}\right)\right]\right| x^{\prime}\right\rangle
=
\sqrt{\frac{m}{2\pi\betavar\hbar^2}}\,
\exp\left[-\frac{\betavar m\omegavar^{\!2}}{2}\left(\frac{x^2+x^{\prime 2}}{2}\right)
-\frac{m}{2\betavar\hbar^2}(x-x^{\prime})^2\right]\,,
\ea
\label{Mehler}
\end{equation}
where
\be
\ba{lll}
\varphi=\omega\beta\hbar\,,\quad&\quad
\betavar=\frac{\sinh\varphi}{\varphi}\beta\,,\quad&\quad
\omegavar=\frac{1}{\cosh(\varphi/2)}\omega\,.
\ea
\label{var3}
\ee
Exploiting ${\bX\cdot\bX}={\bX_{\sigma}\cdot\bX_{\sigma}}$, the $N$-fermion partition function takes the exact form
\be
\!\Tr\!\left[\scalebox{0.95}{$e^{-\beta\left(\frac{1}{2m}{\hbP\cdot\hbP}+\frac{m\omega^2}{2}{\hbX\cdot\hbX}\right)}$}\!\right]\!
=\!\scalebox{0.95}{$\frac{1}{N!}\left(\!\frac{m}{2\pi\betavar\hbar^2}\!\right)^{\!\!\frac{DN}{2}}\!\!\!\!{\dis{\int}}\! D\bX\,$}e^{-\betavar V_{\rm{total}}(\bX)}\,,
\label{HOResult}
\ee
where
\be
V_{\rm{total}}(\bX)=
\frac{1}{2}m\omegavar^2{\bX\cdot\bX}+ V_{\rm{stat}}(\betavar,\bX)\,.
 \label{Vtotal}
\ee
The harmonic trap thus preserves the determinant structure of the statistical potential \textit{exactly}, at every temperature, with the renormalized parameters~(\ref{var3}); in the limit $\omega\to 0$ one recovers $\omegavar\to0$, $\betavar\to\beta$ and the free result~(\ref{ebV}). Since $e^{-\betavar V_{\rm total}(\bX)}\propto\langle\bX|e^{-\beta\hH}|\bX\rangle$, the classical Boltzmann weight of the dual reproduces the full quantum configurational distribution at all temperatures, and in the low-temperature limit
$e^{-\betavar V_{\rm total}(\bX)} \to |\Psi_0(\bX)|^2 e^{-\beta E_0}$: the minima of $V_{\rm total}$ are the maxima of the ground-state probability density.

For a generic $N$-fermion Hamiltonian,
\be
H=\sum_{a=1}^{N}\frac{\vp_a\cdot\vp_{a}}{2m}+V(\vx_{1},\cdots,\vx_{N})
=\frac{\bP\cdot\bP}{2m}+V(\bX)\,,
\label{genericH}
\ee
with permutation-symmetric $V(\bX)=V(\bX_{\sigma})$, a semiclassical expansion~\cite{Wigner1932,Kirkwood1933,FujiwaraOsbornWilk1982} (Methods) gives at leading order
\be
\ba{l}
\quad\Tr\left(e^{-\beta\hH}\right)
\simeq
\scalebox{1.2}{$\frac{1}{\left(\hbar\sqrt{2\pi\beta/m}\right)^{DN}N!}$}{\dis{\int D \bX}}~\sum_{\sigma}\,(-1)^{\sigma}\\
\times\exp\!\left[-\frac{m}{2\beta\hbar^2}\left|\left|{\bX_{\sigma}-\bX}\right|\right|^2
-\beta\dis{\int_{0}^{1}\!\!{\rm{d}}t}~ V\big(\bX+t(\bX_{\sigma}-\bX)\big)\right]\,.
\ea
\label{GeneralResult}
\ee
The classical reduction persists, but each permutation term now carries a trajectory-dependent weight---the potential accumulated along the straight path from $\bX$ to $\bX_\sigma$---so the sum no longer closes into a single determinant (and manifest positivity is no longer guaranteed). Consistency with the exact harmonic result~(\ref{HOResult}) is confirmed at leading order by expanding in small $\varphi$,
\be
\ba{ll}
\frac{1}{\betavar\hbar^2}\simeq
  \frac{1}{\beta\hbar^2}(1-\frac{1}{6}\varphi^2)
 =\frac{1}{\beta\hbar^2}-\frac{\beta\omega^2}{6}\,,\quad&\quad \betavar\omegavar^{\!2}\simeq\beta\omega^2\,.
 \ea
 \label{smallphi}
 \ee

\subsection*{Pauli crystals as global minima of the statistical potential}

Earlier work used pairwise statistical potentials to approximate selected Pauli crystals~\cite{Batle2017,CiftjaBatle2019}. We now apply the exact dual to a two-dimensional harmonic trap, the geometry of the Pauli-crystal experiments~\cite{Holten2021}. We minimize $V_{\rm total}(\bX)$ numerically (Methods) and denote the resulting particle positions by $\vx_a^\star$. The minima exhibit shell structures that coincide with Pauli crystal geometries: the triangular pattern for ${N=3}$ and the ($1{+}5$) shell structure for ${N=6}$ both match the experimentally observed configurations~\cite{Holten2021} (Supplementary Note~1 details the position--momentum duality connecting our configurations to momentum-space imaging). For ${N=55}$, the minimum of $V_{\rm total}$ and the maximum of the ground-state amplitude $|\Psi_0|$ share identical shell structures ($3{+}9{+}9{+}17{+}17$), with a shell-averaged root-mean-square deviation of only $0.004\,a_0$, where $a_0=\sqrt{\hbar/m\omega}$ (Table~\ref{tab:comparison}). A classical, wavefunction-free optimization thus locates the quantum crystal to sub-percent accuracy---and, unlike the ground-state density, it does so at \textit{any} temperature, which is what enables the thermal analysis below.

\begin{table}[!htb]
\centering
\caption{\textbf{$V_{\rm total}$ minimum versus Pauli crystal} (maximum of $|\Psi_{0}|$) for $N=55$ at $\varphi=2$. Shell radii in units of $a_0$. The RMSD is computed over the five shell-averaged radii.}
\label{tab:comparison}
\begin{tabular}{cccc}
\hline\hline
~~~Shell~~~ & ~~~$N_{\rm per\,shell}$~~~ & ~~~$r_{\Psi_0}/a_0$~~~ & ~~~$r_{V_{\rm total}}/a_0$~~~ \\
\hline
1 &  3 & 0.423 & 0.429 \\
2 &  9 & 1.119 & 1.123 \\
3 &  9 & 1.743 & 1.742 \\
4 & 17 & 2.439 & 2.436 \\
5 & 17 & 3.338 & 3.332 \\
\hline
 \multicolumn{4}{c}{Shell-averaged RMSD $= 0.004\,a_0$} \\
\hline\hline
\end{tabular}
\end{table}

\begin{figure}[p]
	\centering
	\includegraphics[width=0.9\columnwidth]{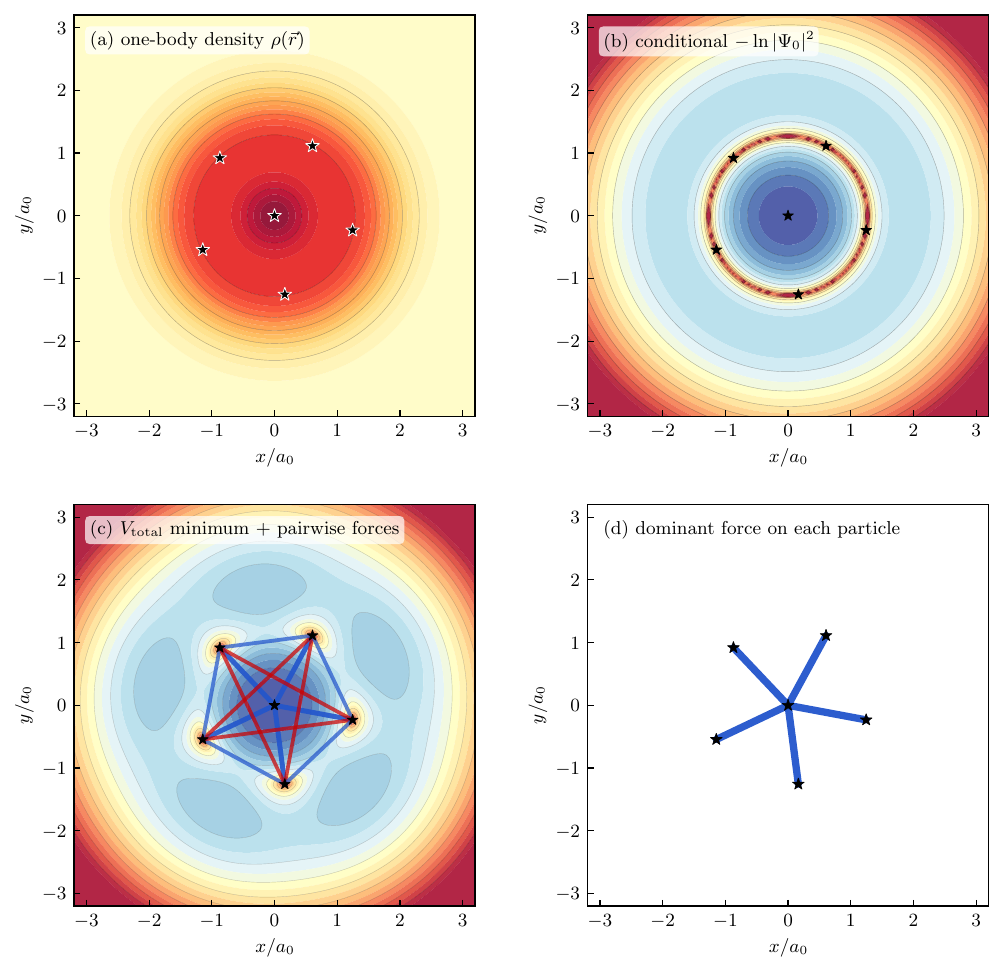}
	\caption{\textbf{The $N=6$ Pauli crystal and its force anatomy} at $\varphi=2$ ($k_{\rm B}T/\hbar\omega=0.5$); lengths in units of $a_0$. Black stars $\star$ mark the minimum of $V_{\rm total}$, i.e.\ the Pauli crystal positions $\vx_a^\star$.
(a)~One-body density $\rho(\vec{r})=\sum_i|\phi_i(\vec{r})|^2$: smooth and rotationally symmetric---the crystal leaves no trace at the single-particle level.
(b)~Conditional density $-\ln|\Psi_0(\vec{r}\,|\,\vx_2^\star,\dots,\vx_N^\star)|^2$ with $N{-}1$ particles fixed, revealing the $1{+}5$ structure as a many-body correlation; cool colors indicate high probability.
(c)~Single-particle slice of the total potential through the minimum. The background contours show $V_{\rm total}(\vec r,\vx_2^\star,\ldots,\vx_N^\star)-V_{\min}$ as the central particle is displaced while the other $N{-}1$ particles remain fixed at their minimizing positions; cool colors indicate lower values. Pairwise statistical forces at the minimum are overlaid: attractive (\red{red}) and repulsive (\blue{blue}), with line thickness proportional to force magnitude. The minimum coincides with the crystal of (b).
(d)~Dominant force on each particle: all five dominant bonds are repulsive (center$\,\leftrightarrow\,$outer)---at $N=6$ no attractive bond dominates (contrast Fig.~\ref{FIG_N55}).}
	\label{FIG_N6}
\end{figure}

\begin{figure}[p]
	\centering
	\includegraphics[width=0.62\columnwidth]{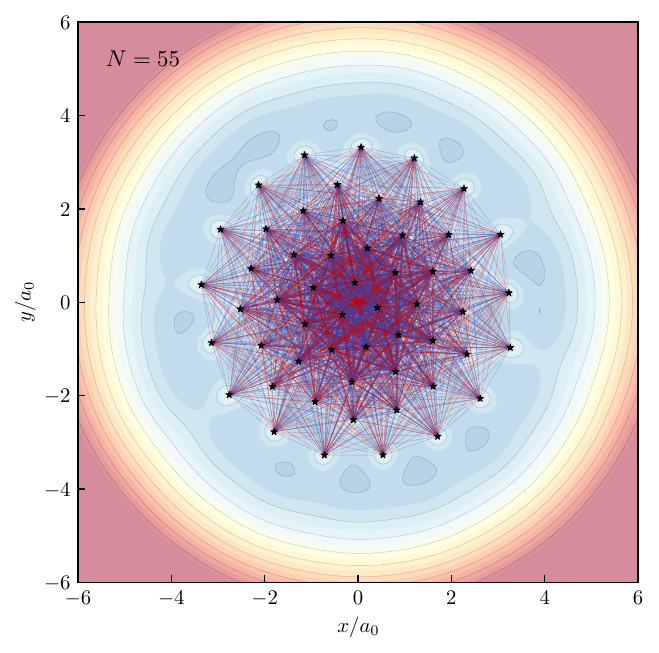}\\[4pt]
	\includegraphics[width=0.62\columnwidth]{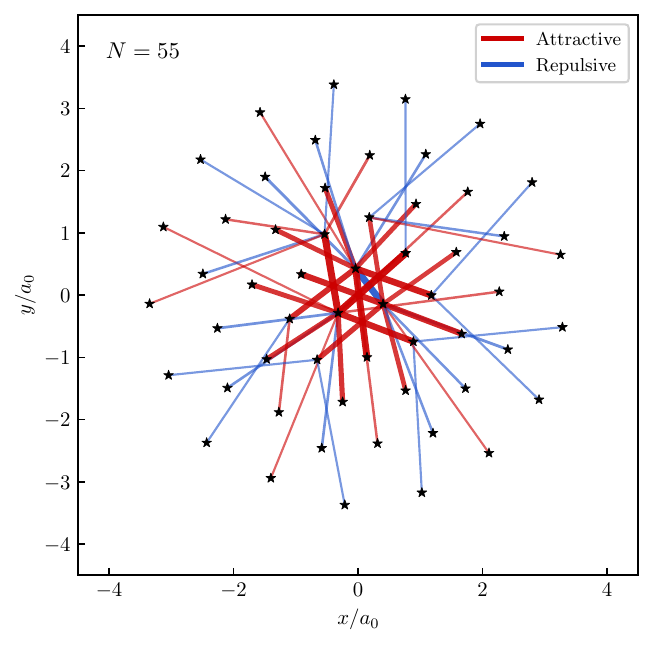}
	\caption{\textbf{Dual force-balance architecture of the $N=55$ Pauli crystal} at $\varphi=2$.
\textbf{Top:} single-particle slice $V_{\rm total}(\vec r,\{\vx_{b\ne a}^\star\})-V_{\min}$ through the minimizing configuration, obtained by displacing the particle closest to the origin while fixing the other 54 particles; cool colors indicate lower values. Pairwise forces at the minimum are overlaid (attractive \red{red}, repulsive \blue{blue}).
\textbf{Bottom:} dominant force on each particle; the displayed network comprises 30 attractive (\red{red}) and 23 repulsive (\blue{blue}) distinct bonds (mutual pairs are drawn as a single bond, hence fewer than $N=55$). The dominant force on inner-shell particles ($3{+}9{+}9$) is attractive, directed toward neighboring shells; on outer-shell particles ($17{+}17$) it is repulsive, directed outward. The globally strongest pairwise force is attractive.}
\label{FIG_N55}
\end{figure}

The force decomposition~(\ref{PairForce}) now supplies what the probability-density description cannot: a mechanical anatomy of the crystal. Pairwise forces evaluated at the minimum reveal, within the canonical decomposition, alternating attractive and repulsive contributions across the shell structure (Figs.~\ref{FIG_N6} and \ref{FIG_N55}). The backgrounds in Fig.~\ref{FIG_N6}(c) and the top panel of Fig.~\ref{FIG_N55} are single-particle slices through this many-body landscape: one particle is displaced while the other $N{-}1$ particles remain fixed at their minimizing positions. For each particle we identify the partner exerting the strongest pairwise force. For $N=6$, the dominant force on every particle is repulsive (center$\,\leftrightarrow\,$outer): no attractive bond dominates. For $N=55$, a qualitatively different picture emerges: the dominant force on inner-shell particles ($3{+}9{+}9$) is \textit{attractive}, directed toward neighboring shells, while that on outer-shell particles ($17{+}17$) is repulsive, directed outward. This dual force-balance architecture---attractive bonds binding the interior, repulsive bonds supporting the exterior---has no counterpart in the purely repulsive $N=2$ statistical potential; it first appears at $N=10$ and becomes prominent at large $N$ (Supplementary Note~3).

\subsection*{Thermal structural transitions and the force crossover}

\begin{figure}[!htb]
\centering
\includegraphics[width=0.95\columnwidth]{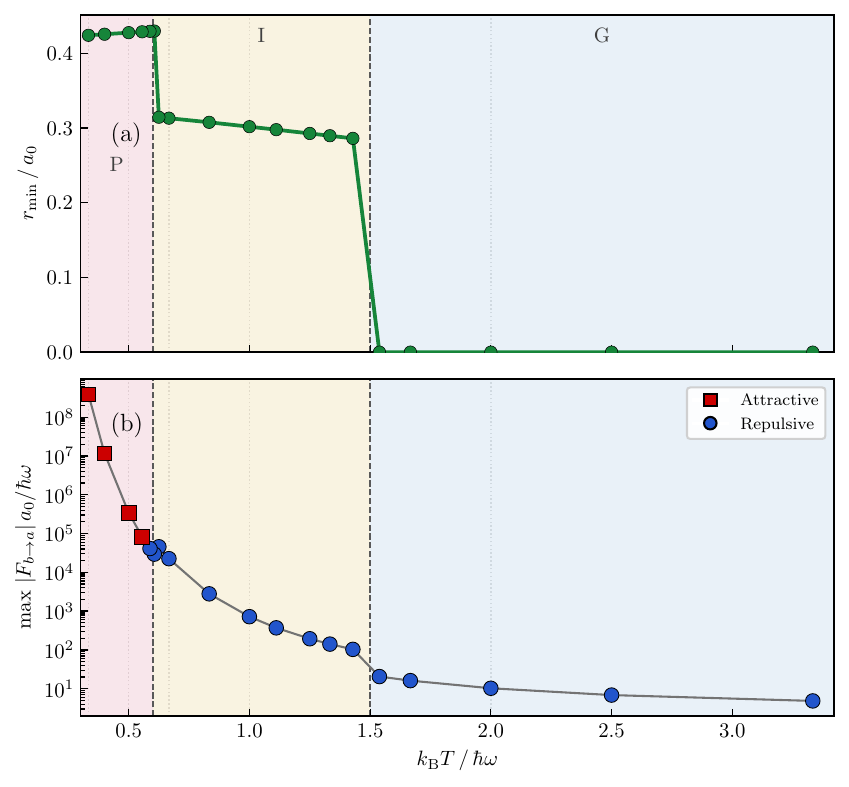}
\caption{\textbf{Melting of the $N=55$ Pauli crystal.} Temperature dependence at the minimum of $V_{\rm total}(\bX)$, with the Pauli-crystal (P), intermediate (I), and gas-like (G) structural regimes shaded across both panels.
(a)~The structural order parameter $r_{\min}=\min_a|\vx_a^\star|$; its discontinuous jumps mark the G$\to$I$\to$P rearrangements.
(b)~Magnitude of the strongest pairwise statistical force, in units of $\hbar\omega/a_0$ and colored by type. The strongest force crosses from repulsive to attractive below $k_{\rm B}T/\hbar\omega\simeq0.6$, coincident with the I$\to$P rearrangement in (a). The force sums and pair counts omitted here, and the $N=6$ null control, are shown in Supplementary Note~4.}
\label{fig:melting}
\end{figure}

Because the mapping is exact at every temperature, heating the crystal is again a classical computation. Figure~\ref{fig:melting} tracks the minimum of $V_{\rm total}$ and its force content for $N=55$ as functions of temperature. Two complementary quantities organize the physics: the structural order parameter $r_{\min}=\min_a|\vx_a^\star|$, the radius of the innermost particle, and the type (attractive or repulsive) of the globally strongest pair force.

Upon cooling, $r_{\min}$ exhibits \textit{discontinuous} jumps separating three structural regimes of the global minimum (Supplementary Note~5): a gas-like phase (G, one particle at the trap center, $r_{\min}\approx 0$), an intermediate phase (I, two innermost particles at $r_{\min}\approx 0.3\,a_0$), and the Pauli crystal (P, three innermost particles forming an equilateral triangle at $r_{\min}\approx 0.43\,a_0$). The G$\to$I and I$\to$P transitions occur at $k_{\rm B}T/\hbar\omega\approx 1.5$ and $\approx 0.6$, respectively. Concurrently, the strongest pair force switches from repulsive to attractive below $k_{\rm B}T/\hbar\omega\simeq 0.6$: the force crossover accompanies the final structural rearrangement into the crystal. For $N=6$, by contrast, the pair topology is temperature-independent and the strongest force remains repulsive at all temperatures (Supplementary Note~4)---at $N=6$ attraction never dominates.

These are not phase transitions in the thermodynamic-limit sense: they are level crossings between competing local minima of the exact effective landscape, first-order-like in character, and they are properties of the distribution's \textit{mode}---the most probable configuration---rather than of its bulk. Direct Monte Carlo sampling of the exact distribution $e^{-\betavar V_{\rm total}}$ (Methods; Supplementary Note~8) shows that the marginal distributions of the first and third radial order statistics remain unimodal and nearly temperature-independent across both transitions: the mode reorganizes discretely while the surrounding thermal mass barely moves. This is consistent with the discrete rearrangements of the innermost particles occurring at strictly constant central pressure, $P(0,T)=\frac{m\omega^2}{2\pi}N$, which holds for any 2D harmonically trapped gas independently of temperature, statistics, or interactions~\cite{ChoKimPark2015} (Supplementary Note~5)---reminiscent of the isobar zigzags of ideal Bose gases~\cite{ChoKimPark2015,ParkKim2010}. The ensemble-level thermal signal appears instead in the bond-sign statistics, to which we now turn.

\begin{figure}[!htb]
\centering
\includegraphics[width=0.8\columnwidth]{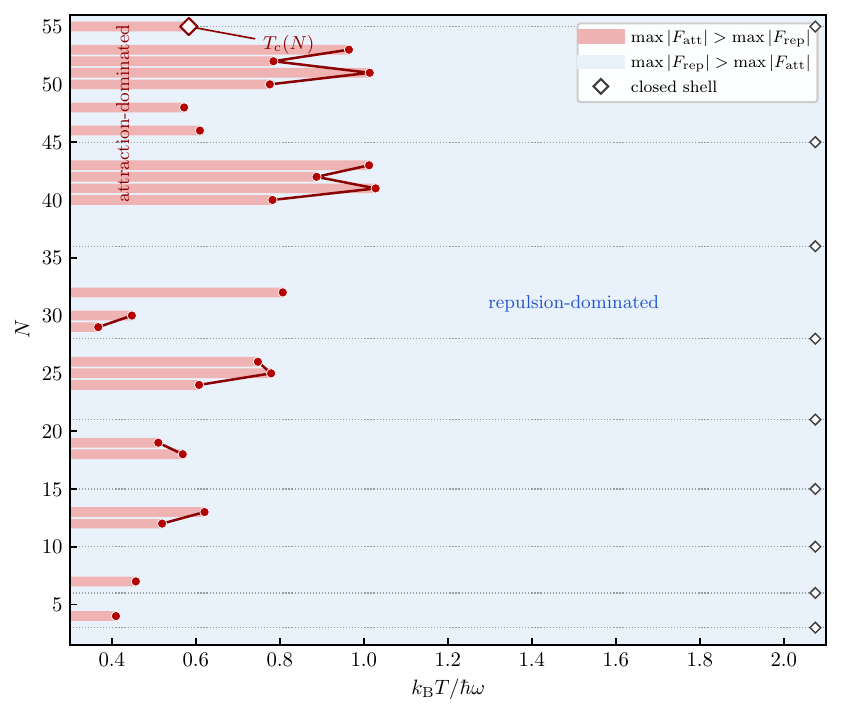}
\caption{\textbf{Force-dominance diagram} in the $(N,\,k_{\rm B}T/\hbar\omega)$ plane for $N=2,\ldots,55$; lower temperatures are on the left. For each discrete $N$, pink shading marks the low-temperature interval where $\max|F_{\rm att}|>\max|F_{\rm rep}|$; the pale-blue background is repulsion-dominated. Red points and connecting segments mark the interpolated crossover boundary $T_{\rm c}(N)$. Filled circles denote open shells; the open diamond at $N=55$ denotes a closed-shell crossover. Horizontal dotted lines and right-edge diamonds mark all closed shells. Closed shells with $N\leq45$ remain repulsion-dominated throughout the sampled range, whereas selected open shells and $N=55$ become attraction-dominated at low temperature.}
\label{fig:phase}
\end{figure}

Whether attraction ever dominates depends on both particle number and temperature (Fig.~\ref{fig:phase}). Closed-shell configurations with $N\leq 45$ remain repulsion-dominated at all temperatures, while some open-shell particle numbers and the largest closed shell studied, $N=55$, develop attractive dominance at low temperature---a shell-filling effect with a finite-$N$ threshold, conceptually reminiscent of the critical particle numbers found for finite ideal Bose gases~\cite{ChoKimPark2015,ParkKim2010}.

\subsection*{Statistical forces are computable shot by shot}

The force decomposition is not merely interpretive: it defines a data-analysis protocol for experiments that already exist. Particle-resolved single-shot imaging in few-body momentum space and continuous real space~\cite{Brandstetter2025,deJongh2025,Holten2021} yields a configuration $\bX$ in each retained realization---for momentum-space imaging, the position--momentum duality of the isotropic harmonic oscillator maps the measured configuration onto $\bX$ (Supplementary Note~1). Given a measured configuration and the known trap frequency and temperature, the matrix $K_{ab}$ of Eq.~(\ref{Kab}) (with $\beta\to\betavar$) and hence \textit{every pairwise force}~(\ref{PairForce}) is computable in closed form---per shot, with no fitting, binning, or histogramming. Each pair is classified as attractive or repulsive by
\be
{\rm sgn}\big(K^{-1\,ab}K_{ab}\big)\;:\;\;+\;\Rightarrow\;{\rm attractive}\,,\qquad -\;\Rightarrow\;{\rm repulsive}\,,
\label{classifier}
\ee
and shot-averaged observables follow: the attractive bond fraction, the identity and type of the strongest bond, and their temperature dependence.

This yields falsifiable predictions, verified here at the ensemble level by direct Monte Carlo sampling of the dual (Methods; Supplementary Note~8). For $N=55$, the probability that the strongest bond of a single shot is attractive rises smoothly from below $10^{-2}$ at $k_{\rm B}T/\hbar\omega=2$ to $0.59$ at $0.4$, crossing one half at $k_{\rm B}T_{\times}/\hbar\omega=0.66(1)$---the ensemble counterpart of the mode-level crossover at ${\approx}0.6$. For $N=6$, the strongest bond remains repulsive in essentially every shot at every temperature ($P<0.07$; a null control). Because the measured shots sample the exact distribution $e^{-\betavar V_{\rm total}}$, the full distribution of bond-sign patterns---not only its mean---is predicted by the same sampling, enabling quantitative comparison at the level of fluctuations. The statistical potential itself is likewise accessible: differences of $V_{\rm total}$ between measured configurations follow from relative configuration densities $\mathcal{P}$, $\Delta V_{\rm total}=-\betavar^{-1}\,\Delta\ln\mathcal{P}$.

\section*{Discussion}

\textit{What the dual is, and is not.}---The statistical potential originates solely from wavefunction symmetry and depends on temperature; as Huang stressed~\cite{Huang1987}, and as debated critically even for $N=2$~\cite{MullinBlaylock}, it is not a true interparticle interaction. Nothing in this work contradicts that view, and no new force of nature is claimed. What we have shown is that the object itself extends beyond $N=2$ into an \textit{exact}, closed-form, positive-weight classical dual of the trapped ideal Fermi gas, valid at all temperatures in harmonic confinement, and that this dual carries canonical mechanical structure---pairwise forces with a definite sign anatomy---whose consequences are the measurable configurational statistics of the gas. Their observable content is the configurational correlations they reproduce: every statement about them can be evaluated on measured single shots via Eqs.~(\ref{PairForce}) and (\ref{classifier}).

\textit{Relation to degeneracy pressure.}---Since the partition functions of the quantum gas and its classical dual are identical, all equilibrium thermodynamics---including degeneracy pressure---is exactly encoded in $V_{\rm stat}$; at $N=2$ this is the classic route by which the statistical potential produces the exchange virial correction~\cite{Uhlenbeck1932,Huang1987,PathriaBeale}. The familiar statement that degeneracy pressure is ``repulsive'' refers to this macroscopic average. The microscopic sign structure is richer: for $N\geq3$ the canonical decomposition contains attractive contributions, organized, at large $N$ and low temperature, into the dual architecture of Fig.~\ref{FIG_N55}---attraction binding inner shells, repulsion supporting outer ones.

\textit{Pauli crystals.}---In the standard description, Pauli crystals are kinematic consequences of antisymmetrization; by construction of the dual, that description and ours are equivalent. Earlier pair-additive models already made selected patterns accessible to approximate classical minimization~\cite{Batle2017,CiftjaBatle2019}. The value of the exact effective-potential formulation is threefold: crystal formation \textit{and} melting become exact classical optimization; it supplies the force anatomy (which bonds hold the crystal, and how they reorganize with $N$ and $T$); and it is computable from experimental data shot by shot. The structural transitions G$\to$I$\to$P, the constancy of the central pressure across them, and the force crossover at $k_{\rm B}T/\hbar\omega\approx0.6$ for $N=55$ are concrete, falsifiable outputs unavailable from the pair-additive or kinematic descriptions~\cite{GajdaEPL2016,Rakshit2017,Xiang2023}.

\textit{Outlook.}---Three extensions appear most immediate. Spinful fermions: antisymmetry then acts jointly on spatial and spin degrees of freedom, restructuring $V_{\rm stat}$. Bosons and other exchange sectors: the determinant is replaced by a permanent, making the two-body attraction the corner of a sign structure that is likewise unexplored for $N\geq3$. Interactions and generic traps: the semiclassical formula~(\ref{GeneralResult}) extends the dual at leading order, with the potential entering through line integrals between permuted configurations---a controlled gateway from the ideal limit toward interacting few-fermion experiments, three-dimensional shell structures, and, more broadly, any regime where antisymmetrization shapes matter, from cold atoms to degenerate astrophysical interiors.

\section*{Methods}

\subsection*{Derivation of the free-fermion mapping}

Let $|\bX\rangle=|\vx_1,\vx_2,\cdots,\vx_{N}\rangle$ be the antisymmetrized position basis state, $|\bX_{\sigma}\rangle=(-1)^{\sigma}|\bX\rangle$, normalized as
\be
\langle \bX|\bX^{\prime}\rangle=\frac{1}{N!}\,\Det\!\Big[\delta^{D}(\vx_{a}-\vx^{\prime}_{b})\Big]_{N\times N}\,.
\label{detX}
\ee
The corresponding momentum states $|\bP\rangle=|\vp_1,\vp_2,\cdots,\vp_{N}\rangle$ obey
\be
\langle\bX|\bP\rangle=\frac{1}{(2\pi\hbar)^{DN/2}\,N!}\,\Det\!\Big[e^{i\vx_{a}\cdot\vp_{b}/\hbar}\Big]_{N\times N}\,.
\label{XP}
\ee
These bases satisfy the completeness relations
\be
\ba{ll}
\dis{\int} D\bX\, |\bX\rangle\langle\bX|={\bf{I}}\,,\quad&\quad\dis{\int} D\bP\, |\bP\rangle\langle\bP|={\bf{I}}\,,
\label{completeness}
\ea
\ee
allowing the partition function to be expressed as a phase-space integral,
\be
\Tr\left(e^{-\frac{\beta}{2m}\hbP\cdot\hbP}\right)=
{\dis{\int} D \bX}{\dis{\int} D \bP}~\langle\bX|\bP\rangle\langle\bP| e^{-\frac{\beta}{2m}\hbP\cdot\hbP}|\bX\rangle\,.
\label{phasespace}
\ee
Substituting Eq.~(\ref{XP}) and performing the Gaussian momentum integration yields Eq.~(\ref{ebV}) with the statistical potential~(\ref{Vstat}).

\subsection*{Mehler kernel}

The Mehler kernel~(\ref{Mehler}) follows from the identities
\be
|x\rangle=e^{-ix\hp/\hbar}|x=0\rangle\,,\quad\qquad
e^{\frac{i}{2\hbar}(\ln\cosh\varphi)\,(\hat x\hat p+\hat p\hat x)}|x\rangle=\frac{1}{\sqrt{\cosh\varphi}}|x/\cosh\varphi\rangle\,,
\label{mehlerids}
\ee
together with the decomposition of the Boltzmann factor,
\begin{equation}
\begin{aligned}
\exp\!\left[
-\beta\left(\frac{\hat p^{\,2}}{2m}
+\frac{m\omega^{2}}{2}\hat x^{\,2}\right)
\right]
&=
\exp\!\left(
-\beta\,\frac{\tanh\varphi}{2m\varphi}\,\hat p^{\,2}
\right)
\exp\!\left[
\frac{i}{2\hbar}(\ln\cosh\varphi)\,
(\hat x\hat p+\hat p\hat x)
\right]\\
&\quad\times
\exp\!\left(
-\beta\,\frac{m\omega^2\tanh\varphi}{2\varphi}\,\hat x^{\,2}
 \right)\,,
\end{aligned}
\label{mehlerdecomp}
\end{equation}
where $\varphi=\omega\beta\hbar$. Applying the $N$-particle, $D$-dimensional version of Eq.~(\ref{Mehler}) between the antisymmetrized bases~(\ref{detX})--(\ref{completeness}), and using ${\bX\cdot\bX}={\bX_{\sigma}\cdot\bX_{\sigma}}$ to factor the harmonic weight out of the permutation sum, gives Eq.~(\ref{HOResult}).

\subsection*{Semiclassical expansion for generic potentials}

We derive Eq.~(\ref{GeneralResult}) by a leading-order semiclassical expansion of the thermal kernel; the potential contributes through a line integral along the straight path connecting $\bX$ and $\bX_\sigma$.

We employ the Baker--Campbell--Hausdorff (BCH) variant proved below,
\be
\ln\left(e^{\hA}e^{-\hA-\hB}e^{\hB}\right)=
\int_{0}^{1}{\rm{d}}t~\left(1-e^{t\,{\ad }\hA}\right)\hB+\cO(\hB^2)\,,
\label{BCH}
\ee
with $\hat A=\frac{\beta}{2m}\hat{\bP}^{\,2}$, $\hat B=\beta V(\hat{\bX})$, and $({\ad}{\hat{A}})(\hat{B}) = [\hat{A}, \hat{B}]$.

For a single particle in one dimension (${N=1}$, ${D=1}$), the BCH expansion~(\ref{BCH}) implies
\be
e^{-\beta\hH}\simeq e^{-\frac{\beta}{2m}\hp^2}\!\exp\!\left[\beta\!{\int_{0}^{1}\!\!{\rm{d}}t\left(1-e^{t\frac{\beta}{2m}{\ad \,}\hp^2}\right)\!V(\hx)}\right]\!e^{-\beta V(\hx)}\,.
\label{singleH}
\ee
Inserting a completeness relation, we note
\be
\langle x|e^{-\beta\hH}|x^{\prime}\rangle\simeq\dis{\int}\frac{\rd p}{\sqrt{2\pi\hbar}}~ e^{-\frac{\beta}{2m}p^2+ixp/\hbar-\beta V(x^{\prime})}
\langle p|\exp\!\left[\beta\!\dis{\int_{0}^{1}{\rm{d}}t}\left(1-e^{t\frac{\beta}{2m}{\ad \,}\hp^2}\right)\!V(\hx)\right]\!|x^{\prime}\rangle\,.
\label{intp}
\ee
To leading order, the Gaussian integration effectively replaces $\hbar p$ by a derivative acting on $x$, implying $\hbar p = \cO(1)$ in the semiclassical scaling: for a sufficiently smooth function $F$,
\be
\sqrt{\frac{\beta}{2m\pi}}{\dis{\int}}\!\rmd p~ e^{-\frac{\beta }{2m}p^{2} +ixp/\hbar}F\big(-i\textstyle{\frac{\hbar\beta}{m} }p\big)
=F\big(-\frac{\beta\hbar^2}{m}\partial_{x}\big)e^{-\frac{m}{2\hbar^2\beta}x^{2}}=e^{-\frac{m}{2\beta\hbar^2}x^{2}}\big[F(x)+\cO(\hbar^2)\big]\,.
\label{elementary}
\ee
Accordingly, the adjoint action of the kinetic operator yields an $\cO(1)$ contribution and must be retained:
\be
(\ad\,\hp^2) {V}(\hx)=\! -2i\hbar\hp\,\partial_{x}V(\hx)+\hbar^2\partial_{x}^2V(\hx)\simeq \!  -2i\hbar\hp\,\partial_{x}V(\hx)\,,
\label{adaction}
\ee
which generates a translation of the argument of the potential,
\be
e^{t\frac{\beta}{2m}\ad \hat p^{\,2}}V(\hat x)\simeq e^{ -it\frac{\beta}{m}\hbar \hp\,\partial_{x}}V(x)=V\big(x-\textstyle{it\frac{\beta}{m}\hbar} \hp\big)\,.
\label{translation}
\ee
This translation property is the origin of the trajectory-dependent structure of Eq.~(\ref{GeneralResult}). Restricting to terms linear in $V$ isolates the leading semiclassical behavior:
\be
\begin{aligned}
\langle x|e^{-\beta\hH}|x^{\prime}\rangle
&\simeq{\!\dis{\int}}\frac{\rd p}{{2\pi\hbar}}\,
e^{-\frac{\beta}{2m}p^2+\frac{i}{\hbar}(x-x^{\prime})p
-\beta{\scalebox{1}{$\int_{0}^{1}$}{\!\rm{d}}t\,} V(x^{\prime}-it\frac{\beta}{m}\hbar p)}\\
&=\sqrt{\frac{m}{2\pi\beta\hbar^2}}\,
e^{-\frac{m}{2\beta\hbar^2}(x-x^{\prime})^2
-\beta\scalebox{1}{$\int_{0}^{1}$}{\rm{d}}t\, V\left(x^{\prime}+t(x-x^{\prime})\right)}\,,
\end{aligned}
\label{intp2}
\ee
in agreement with Ref.~\cite{FujiwaraOsbornWilk1982}: at this order the kernel samples the potential along the straight segment connecting $x^{\prime}$ and $x$.

The single-particle expressions~(\ref{singleH})--(\ref{intp2}) generalize straightforwardly to $N$ fermions in $D$ dimensions by replacing $x,p$ with $\bX,\bP$. Using Eqs.~(\ref{XP}) and (\ref{completeness}) and summing over permutations $\sigma$, the partition function becomes
\be
\begin{aligned}
\Tr\left(e^{-\beta\hH}\right)
&\simeq
\frac{1}{N!\left(\hbar\sqrt{2\pi\beta/m}\right)^{DN}}
{\dis{\int D \bX}}~\sum_{\sigma}\,(-1)^{\sigma}\\
&\quad\times
\exp\!\left[-\frac{m}{2\beta\hbar^2}\left|\left|{\bX_{\sigma}-\bX}\right|\right|^2
-\beta\dis{\int_{0}^{1}\!\!{\rm{d}}t}~ V\big(\bX+t(\bX_{\sigma}-\bX)\big)\right]\,,
\end{aligned}
\label{Nsemi}
\ee
which reproduces Eq.~(\ref{GeneralResult}). For a classical potential of the form
\be
V(\bX)=\sum_{a}~V_{\rm{ext}}(\vx_{a})+\sum_{a<b}~V_{\rm{int}}(\vx_{a}-\vx_{b})\,,
\label{Vclass}
\ee
the semiclassical expression generates effective couplings,
\be
V\big(\bX+t(\bX_{\sigma}-\bX)\big)=\dis{
\sum_{a}}~V_{\rm{ext}}\big((1-t)\vx_{a}+t\vx_{\sigma(a)}\big)
+\dis{\sum_{a<b}}~V_{\rm{int}}\big((1-t)(\vx_{a}-\vx_{b})+t(\vx_{\sigma(a)}-\vx_{\sigma(b)})\big)\,,
\label{effcouplings}
\ee
including effective many-body contributions that cannot, in general, be reduced to pairwise interactions.

\subsection*{A BCH identity}

We establish the variant of the BCH formula used above:
\be
\begin{aligned}
\ln\left(e^{A}e^{-A-B}e^{B}\right)
&=B-\int_{0}^{1}{\rm{d}}t~\sum_{n=1}^{\infty}\frac{1}{n}
\left(1-e^{t{\ad }A}e^{-t{\ad }(A+B)}e^{{\ad }B}\right)^{n-1}e^{t{\ad }A}B\\
&=
-\frac{1}{2}\left[A,B\right]
+\frac{1}{6}\Big(\left[B,\left[A,B\right]\right]-\left[A,\left[A,B\right]\right]\Big)\\
&\quad+\scalebox{0.9}{$\displaystyle
\frac{1}{24}\Big(\left[B,\left[A,\left[A,B\right]\right]\right]
-\left[A,\left[A,\left[A,B\right]\right]\right]
-\left[B,\left[B,\left[A,B\right]\right]\right]\Big)
+\cO(5)\,.$}
\end{aligned}
\label{BCHsm}
\ee
In particular, retaining terms linear in $B$ gives Eq.~(\ref{BCH}):
\be
\ln\left(e^{A}e^{-A-B}e^{B}\right)=
\int_{0}^{1}{\rm{d}}t~\left(1-e^{t{\ad }A}\right)B+\cO(B^2)
=-\sum_{n=1}^{\infty}
\frac{(\ad A)^{n}}{(n+1)!}B+\cO(B^2)\,.
\label{BCHlinear}
\ee

\textit{Proof.} Introduce a real parameter $t \in \mathbb{R}$ and define
\begin{equation}
C(t) := \ln\!\Big(
e^{tA} e^{-t(A+B)}e^{B}
\Big),
\qquad
e^{C(t)} =
e^{tA} e^{-t(A+B)}e^{B}\,,
\end{equation}
so that $C(0)=B$ and
\begin{equation}
\frac{\rmd}{\rmd t} e^{C(t)}
=-e^{tA}Be^{-t(A+B)}e^{B}=-e^{tA}Be^{-tA}e^{C(t)}\,.
\end{equation}
Define
\begin{equation}
F(s,t) := \big[\partial_t e^{sC(t)}\big]e^{-sC(t)}\,,
\end{equation}
which satisfies $F(0,t)=0$. Using
\begin{equation}
e^{\ad C(t)}
=e^{t\ad A} e^{-t\ad(A+B)}e^{\ad B}\,,
\end{equation}
we obtain
\be
F(1,t)=-e^{t\ad A}B\,.
\ee
Furthermore,
\begin{equation}
\partial_s F(s,t)
=
e^{s\ad C(t)}\!\left[\frac{dC(t)}{dt}\right]\,.
\end{equation}
Integrating over $s$ from $0$ to $1$,
\begin{equation}
F(1,t)
=
\int_0^1 \rmd s\, e^{s\ad C(t)}\!\left[\frac{\rmd C(t)}{\rmd t}\right]
=
\sum_{n=0}^\infty
\frac{(\ad C(t))^n}{(n+1)!}
\left[\frac{\rmd C(t)}{\rmd t}\right]
= G\big(\ad C(t)\big)\!\left[\frac{\rmd C(t)}{\rmd t}\right]\,,
\end{equation}
where
\begin{equation}
G(x)=\frac{e^x-1}{x}
=\sum_{n=0}^\infty \frac{x^n}{(n+1)!}\,,\qquad
G(x)^{-1}
=\frac{x}{e^x-1}
=\sum_{n=1}^\infty \frac{(1-e^x)^{n-1}}{n}\,.
\end{equation}
Thus
\begin{equation}
\frac{\rmd C(t)}{\rmd t}
=
\sum_{n=1}^\infty
\frac{1}{n}
\big(1 - e^{\ad C(t)}\big)^{n-1}
F(1,t)\,.
\end{equation}
Integrating from $t=0$ to $1$ yields Eq.~(\ref{BCHsm}). All series are understood as formal power series. \hfill$\square$

\subsection*{Numerical optimization}

All minimizations of $V_{\rm total}(\bX)$ are performed using the L-BFGS-B algorithm with analytic gradients. Because $V_{\rm total}$ is a nonconvex function of $DN=2N$ variables (110 variables for $N=55$), we employ a multi-start strategy: for each $(N,\varphi)$, the optimizer is run from at least 300 independent random initial configurations using shell-type seeds (concentric rings with random perturbations), and the lowest-energy result is retained. For $N=55$ at $\varphi=2$, the best minimum is reproduced across multiple seeds to within $\sim 10^{-3}\,a_0$ in particle positions, supporting convergence to a low-lying minimum. The force-dominance diagram (Fig.~\ref{fig:phase}) uses 1000 seeds per $(N,\varphi)$ point. The maximization of $|\Psi_0(\bX)|^2$ for the ground-state Slater determinant follows an analogous multi-start L-BFGS-B procedure with 300 seeds and numerical gradients.

\subsection*{Ensemble Monte Carlo}

The exact configurational distribution $e^{-\betavar V_{\rm total}}$ is sampled directly, exploiting its strictly positive weight. We use single-particle Metropolis moves, with step sizes adapted during burn-in only, combined with replica exchange across the temperature ladder. Because $V_{\rm total}$ depends on temperature through $(\betavar,\omegavar)$, swaps between neighboring temperatures $i,j$ holding configurations $\bX_i,\bX_j$ are accepted with probability $\min\{1,\,e^{L_i(\bX_j)+L_j(\bX_i)-L_i(\bX_i)-L_j(\bX_j)}\}$, where $L_i(\bX)=\ln\Det K(\betavar^{(i)},\bX)-\tfrac{1}{2}\betavar^{(i)}\omegavar^{(i)2}\,\bX\cdot\bX$ is the log-weight at temperature $i$. For $N=55$ we use 25 temperatures ($0.40\le k_{\rm B}T/\hbar\omega\le 2.00$, dense near the structural transitions), $3\times10^{4}$ measurement sweeps per temperature after burn-in, and three independent runs initialized from gas-like, crystalline, and mixed configurations; all temperatures reach split-chain $\hat R\le1.001$ with effective sample sizes $\gtrsim8{,}000$. The sampler was validated against the exact $N=2$ separation distribution (Kolmogorov--Smirnov distance $0.013$--$0.018$, inside the $95\%$ band at the measured effective sample size) and reproduces the exact absence of attractive pairs at $N=2$. Per-sample bond classification uses Eq.~(\ref{classifier}); the ensemble crossover temperature is obtained by interpolating $P(\text{strongest bond attractive})=\tfrac12$, with binomial (Wilson) intervals computed on the sign-chain effective sample size and a run-to-run spread of $0.003$ (Supplementary Note~8).

\subsection*{Use of generative AI}

Claude (Anthropic) assisted with code development. Codex (OpenAI) assisted with code review, reference checking, and manuscript editing. The authors independently verified all analytical derivations, numerical outputs, source citations, and final text and take full responsibility for the work. No generative AI was used to create research data or figures.

\subsection*{Data availability}

The numerical data supporting the figures are generated by the publicly available code below. Cached data for the computationally intensive temperature and particle-number scans, together with the raw outputs of the independent verification calculations, are included in the public repository.

\subsection*{Code availability}

All custom code used for the numerical simulations and figure generation is publicly available at \url{https://github.com/Jeong24th/statistical_potential}~\cite{code}. The version accompanying this submission is archived as commit \href{https://github.com/Jeong24th/statistical_potential/commit/6ffe4b8b40e7d3473f5023735d438d468c668059}{6ffe4b8}.



\subsection*{Acknowledgements}
We thank Nayeon Kim for early discussions in 2013 that first revealed attraction at ${N=3}$. This work was supported by the National Research Foundation of Korea (NRF) through grant RS-2023-NR077094 and through the Center for Quantum Spacetime (CQUeST), grant RS-2020-NR049598.

\subsection*{Author contributions}
K.L. and S.O. carried out the analytical derivations and numerical computations and prepared the figures. Y.W.C. provided overall scientific review. J.-H.P. conceived and supervised the project. All authors reviewed and approved the manuscript.

\subsection*{Competing interests}
The authors declare no competing interests.

\clearpage
\setcounter{figure}{0}
\setcounter{table}{0}
\setcounter{equation}{0}
\renewcommand{\thefigure}{S\arabic{figure}}
\renewcommand{\thetable}{S\arabic{table}}
\renewcommand{\theequation}{S\arabic{equation}}
\renewcommand{\theHfigure}{S\arabic{figure}}
\renewcommand{\theHtable}{S\arabic{table}}
\renewcommand{\theHequation}{S\arabic{equation}}
\begin{center}
{\LARGE\bfseries Supplementary Information}\\[10pt]
{\normalsize for ``Attractive statistical forces and Pauli crystal formation in trapped Fermi gases''}
\end{center}
\vspace{6pt}
\section*{Supplementary Note 1: Experiment and\\ position--momentum duality}
\label{SI:exp}

The minima of $V_{\rm total}(\bX)$ are consistent with experimentally observed Pauli crystals. Holten \textit{et al.}\ [main text Ref.~16] measured the configuration probability densities of $N=3$ and $N=6$ noninteracting fermions in a 2D harmonic trap. In these experiments, each single-shot measurement yields the momenta of all $N$ particles; the configuration density is then constructed by rotating each shot to a common symmetry axis following Rakshit \textit{et al.}\ [main text Ref.~12]. Although the measurements are performed in momentum space, the 2D isotropic harmonic oscillator is invariant under
\be
\frac{x}{\sqrt{\hbar/m\omega}}\quad\Longleftrightarrow\quad \frac{p}{\sqrt{\hbar m\omega}}\,,
\qquad\qquad
V_{\rm{stat}}(\betavar,\bX)\quad\Longleftrightarrow\quad V_{\rm{stat}}(\betavar,\bP/m\omega)\,,
\label{xpduality}
\ee
so the Pauli crystal geometry---and every force-level statement of the main text---is identical in position and momentum space up to the scale factor $m\omega$. This duality is what allows the shot-by-shot force protocol of the main text to be applied directly to momentum-space single-shot data.

For $N=3$, the experimentally observed configuration density displays a triangular pattern (cf.\ Supplementary Fig.~\ref{fig:SM_multiN_forces}), and for $N=6$ it exhibits a $1{+}5$ shell structure (main text Fig.~2)---both matching the minima of $V_{\rm total}(\bX)$. Our prediction for $N=55$ (main text Fig.~3) awaits experimental verification.

\section*{Supplementary Note 2: One-body vs.\ many-body density}
\label{SI:density}

\begin{figure}[H]
\centering
\includegraphics[width=0.48\textwidth]{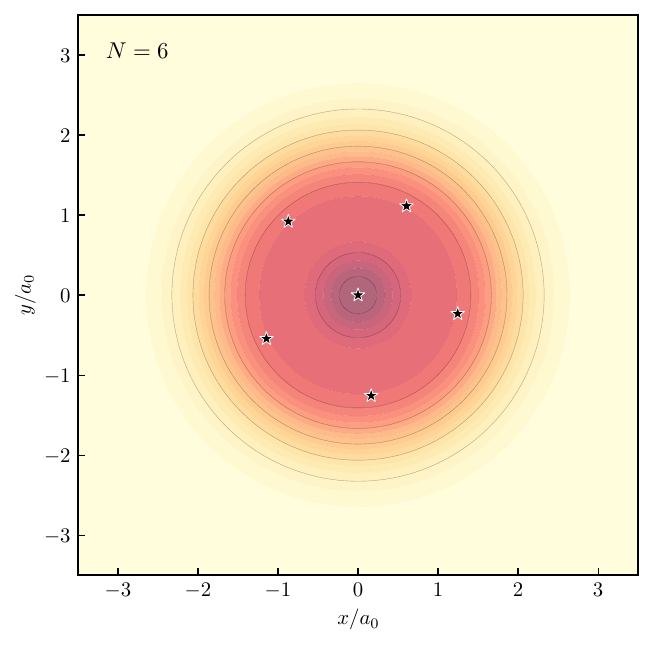}%
\hfill
\includegraphics[width=0.48\textwidth]{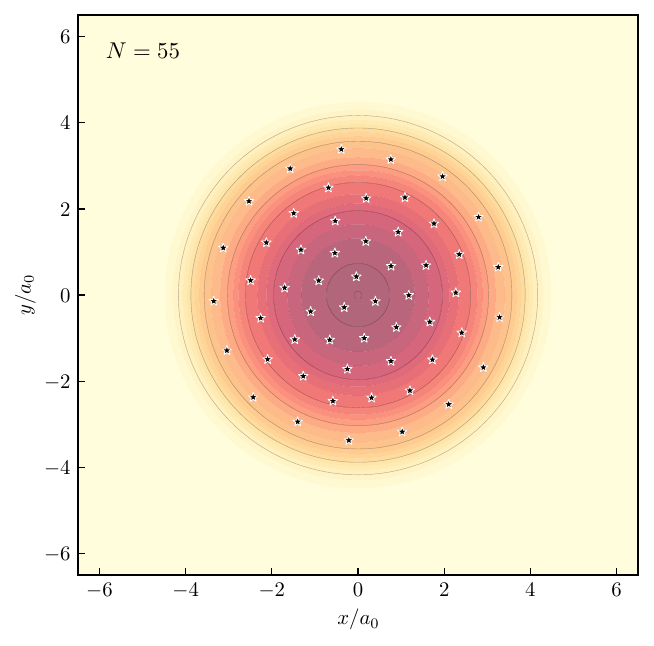}
\caption{One-body density $\rho(\vec{r})$ for $N=6$ (left) and $N=55$ (right). The density is smooth and rotationally symmetric. Black stars $\star$ mark the Pauli crystal positions (minimum of $V_{\rm total}$), which leave no trace in $\rho$.}
\label{fig:SM_1body}
\end{figure}

\begin{figure}[H]
\centering
\includegraphics[width=0.48\textwidth]{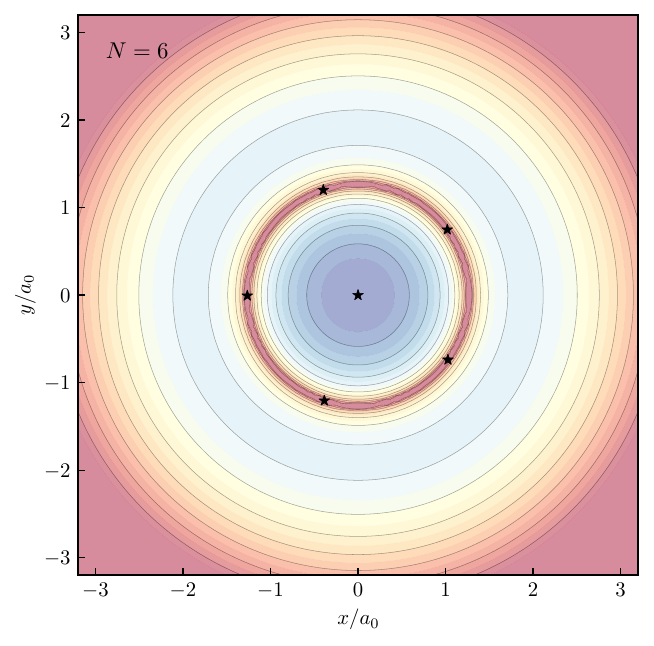}%
\hfill
\includegraphics[width=0.48\textwidth]{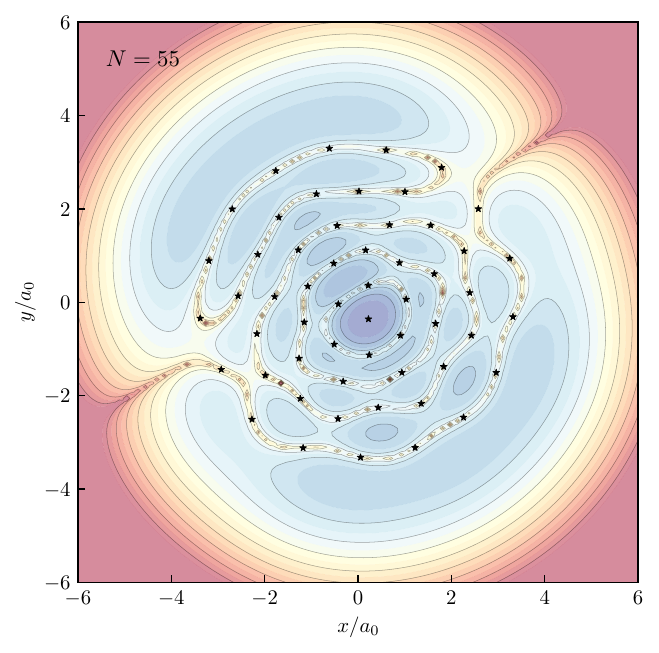}
\caption{Conditional density $-\ln|\Psi_0|^2$ for $N=6$ (left) and $N=55$ (right), with $N{-}1$ particles fixed at the Pauli crystal configuration ($|\Psi_0|$ maximum). Cool colors indicate high probability. The minima coincide with those of $V_{\rm total}$ (main text Figs.~2, 3 and Table~1). For $N=55$, swirl-like features trace nodal lines where $\Psi_0=0$.}
\label{fig:SM_conditional}
\end{figure}

The one-body density $\rho(\vec{r})=\sum_{i}|\phi_i(\vec{r})|^2$ (Supplementary Fig.~\ref{fig:SM_1body}) is smooth and featureless---crystal structure is invisible at the single-particle level. In contrast, the conditional $N$-body density (Supplementary Fig.~\ref{fig:SM_conditional}) reveals the crystal, confirming that Pauli crystal order is a genuine many-body correlation.

\FloatBarrier
\section*{Supplementary Note 3: Particle-number dependence}
\label{SI:Ndep}

\begin{figure}[H]
\centering
\includegraphics[width=0.9\textwidth]{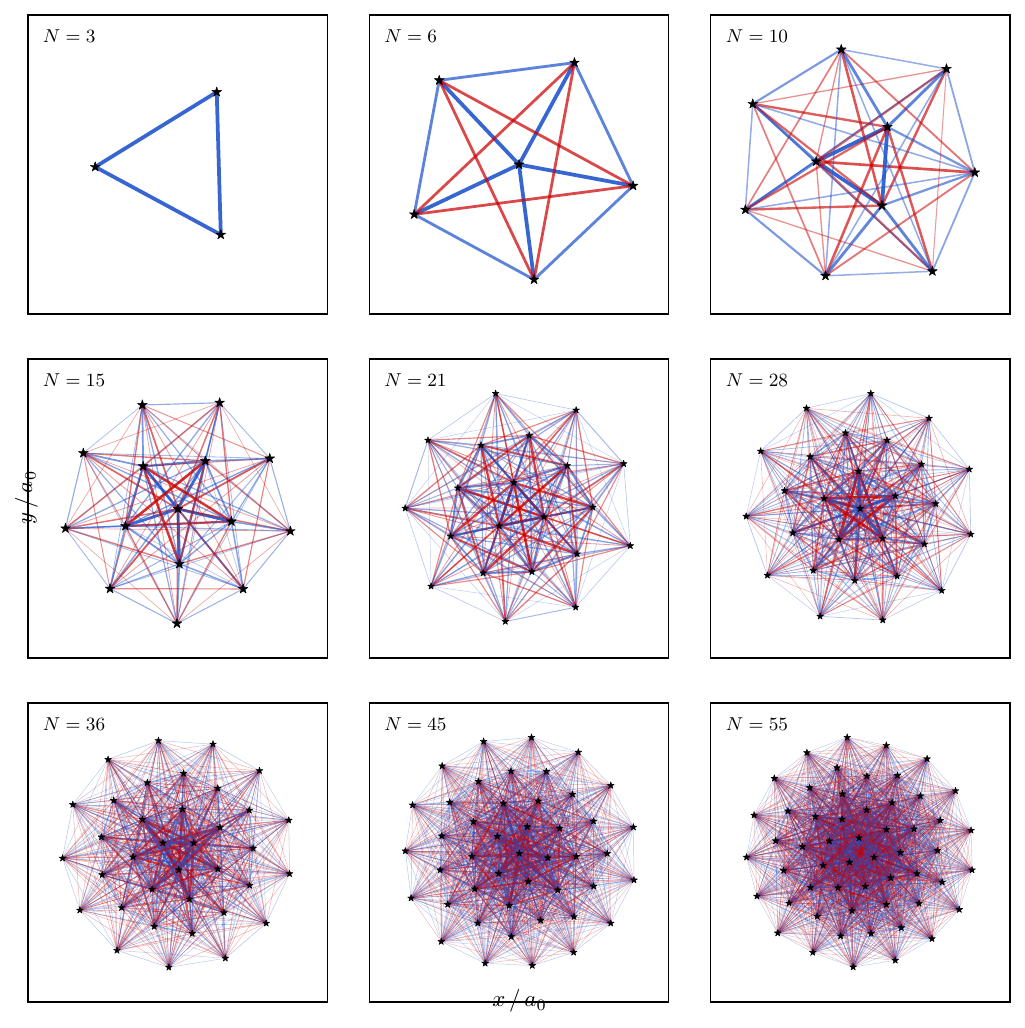}
\caption{Pairwise statistical forces at the $V_{\rm total}$ minimum for closed-shell particle numbers $N=3$--$55$ at $\varphi=2$. Attractive (\red{red}) and repulsive (\blue{blue}); line thickness proportional to force magnitude.}
\label{fig:SM_multiN_forces}
\end{figure}

Supplementary Fig.~\ref{fig:SM_multiN_forces} shows the pairwise statistical forces at the minimum of $V_{\rm total}(\bX)$ for all closed-shell particle numbers from $N=3$ to $N=55$ at $\varphi=2$. Attractive (red) and repulsive (blue) contributions are present for all $N\geq 6$, with the network of force lines becoming increasingly complex with $N$.

\begin{figure}[H]
\centering
\includegraphics[width=0.9\textwidth]{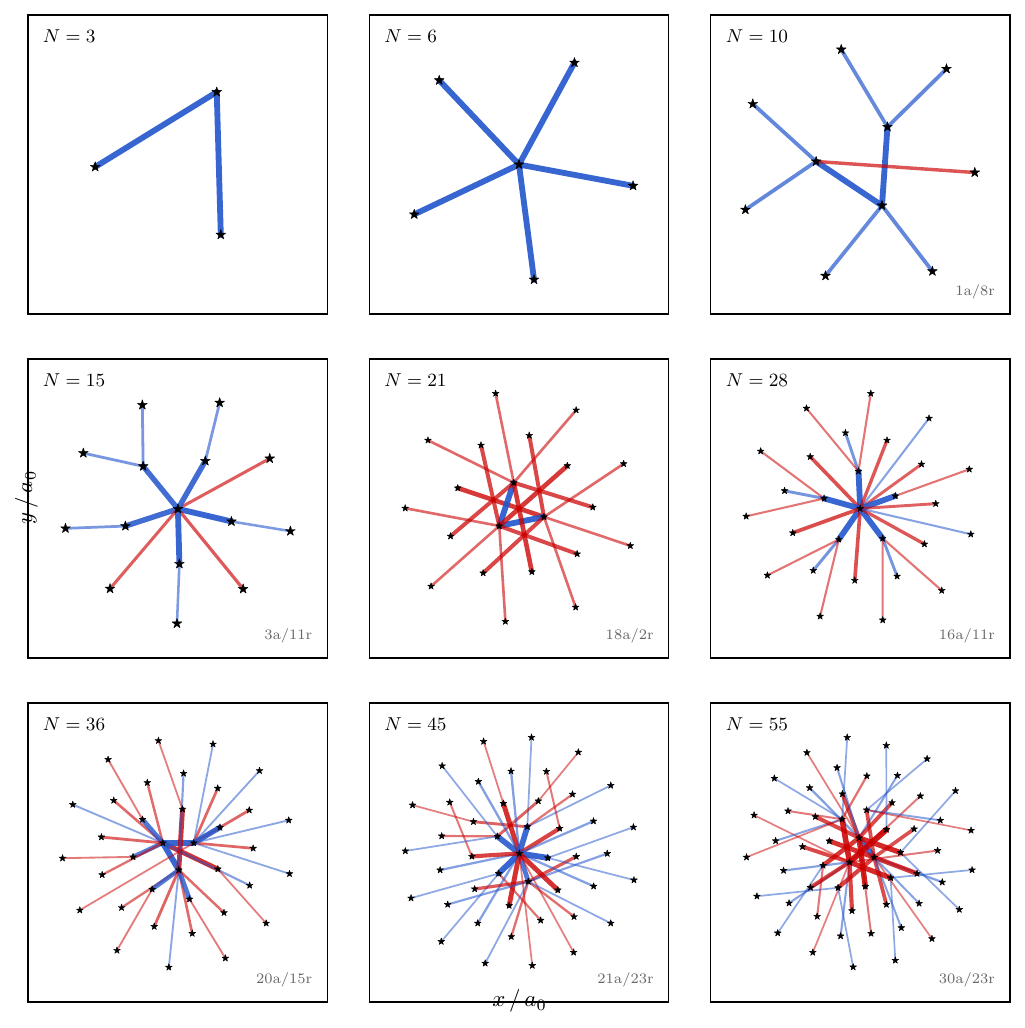}
\caption{Dominant pairwise force on each particle for $N=3$--$55$ at $\varphi=2$. Each line connects a particle to the partner exerting the strongest force. Labels indicate the number of attractive~(a) and repulsive~(r) bonds. The first attractive dominant bond appears at $N=10$ (1a/8r), marking the onset of the dual force-balance architecture that becomes prominent at large $N$ (cf.\ main text Fig.~3).}
\label{fig:SM_multiN_strongest}
\end{figure}

Supplementary Fig.~\ref{fig:SM_multiN_strongest} displays the dominant pairwise force on each particle for the same set of $N$ values. For $N=3$ and $N=6$, all dominant forces are repulsive. Attractive dominant bonds first appear at $N=10$ and become increasingly prominent at larger $N$, marking the onset of the dual force-balance architecture described in the main text.

\begin{figure}[H]
\centering
\includegraphics[width=0.80\textwidth]{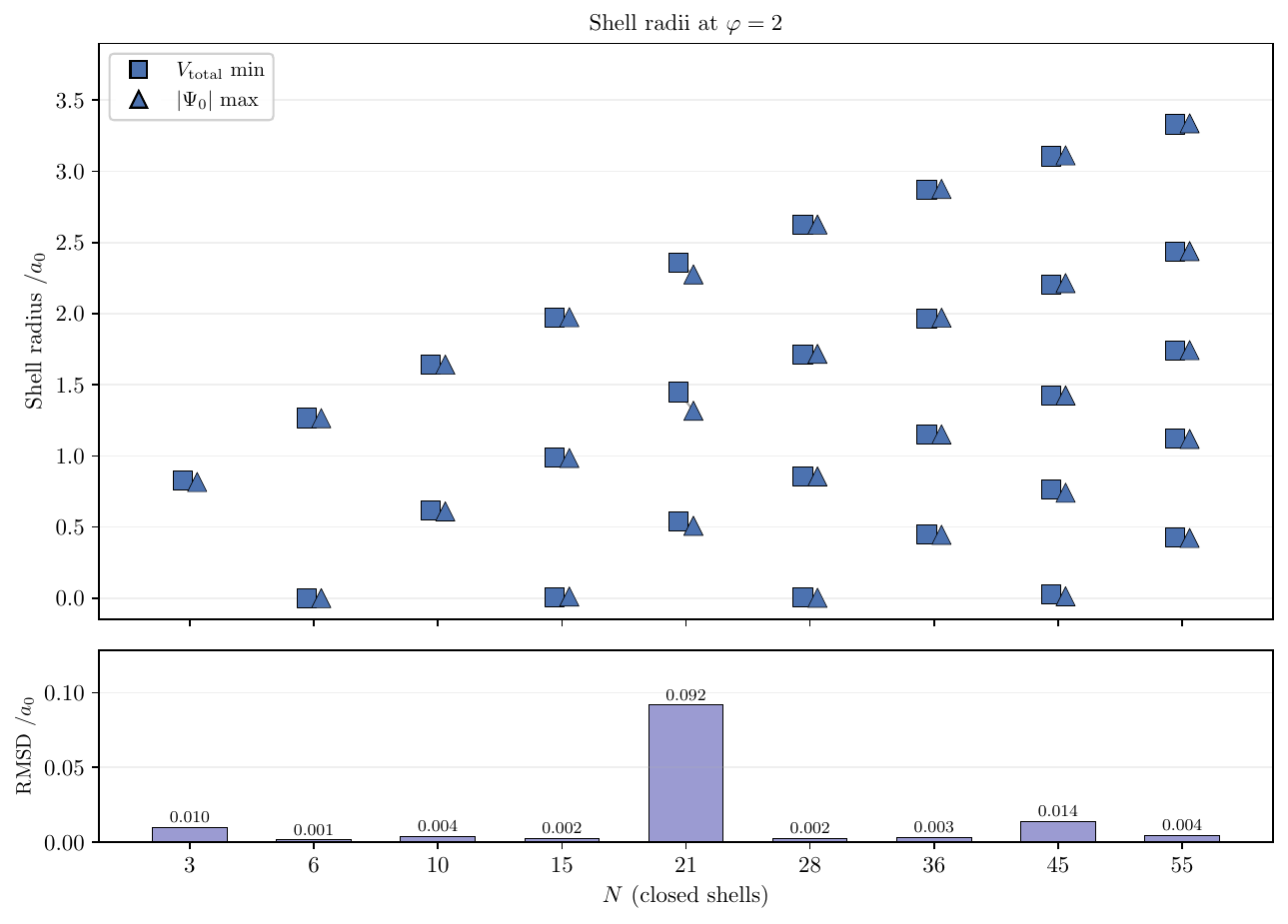}
\caption{Shell radii at the $V_{\rm total}$ minimum (squares) and the $|\Psi_0|$ maximum (triangles) for closed-shell $N$ at $\varphi=2$; the y-axis position directly encodes the shell index (inner~$\to$~outer). The two configurations yield essentially identical shell radii across all $N$ (RMSD~$\leq 0.014\,a_0$), except for $N=21$, where competing near-degenerate shell distributions ($3{+}9{+}9$ for $V_{\rm total}$ vs.\ $3{+}7{+}11$ for $|\Psi_0|$) produce a larger deviation.}
\label{fig:SM_shell_radii}
\end{figure}

Supplementary Fig.~\ref{fig:SM_shell_radii} compares the shell radii extracted from the minimum of $V_{\rm total}$ with those obtained from the maximum of $|\Psi_0|$ for all closed-shell particle numbers $N$. The two sets of radii are nearly indistinguishable, confirming that the statistical potential faithfully reproduces the Pauli-crystal geometry across system sizes.

\FloatBarrier
\section*{Supplementary Note 4: Temperature dependence}
\label{SI:temp}

\begin{figure}[H]
\centering
\includegraphics[width=0.95\textwidth]{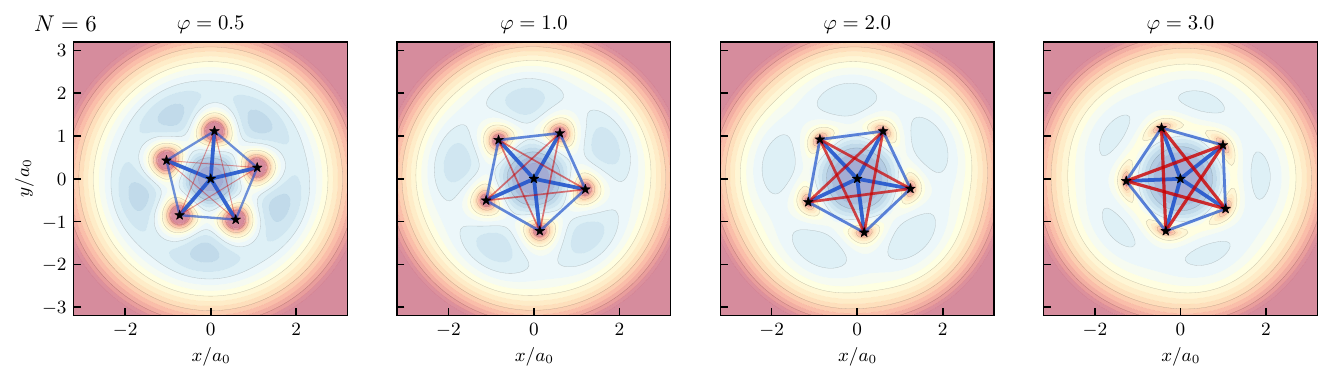}
\caption{$V_{\rm total}$ landscape and pairwise forces for $N=6$ at $\varphi=0.5$, $1.0$, $2.0$, and $3.0$ (left to right). The 5-fold crystal structure sharpens with decreasing temperature.}
\label{fig:SM_temp_N6}
\end{figure}

Supplementary Fig.~\ref{fig:SM_temp_N6} shows the $V_{\rm total}$ landscape and pairwise forces for $N=6$ at four temperatures ($\varphi=\hbar\omega/\kB T=0.5$, $1$, $2$, $3$). At high temperature ($\varphi=0.5$), the contour is nearly circular (trap-dominated) and the forces are weak. As the temperature decreases, the 5-fold structure sharpens and the attractive contributions (red) become visible.

\begin{figure}[H]
\centering
\includegraphics[width=0.95\textwidth]{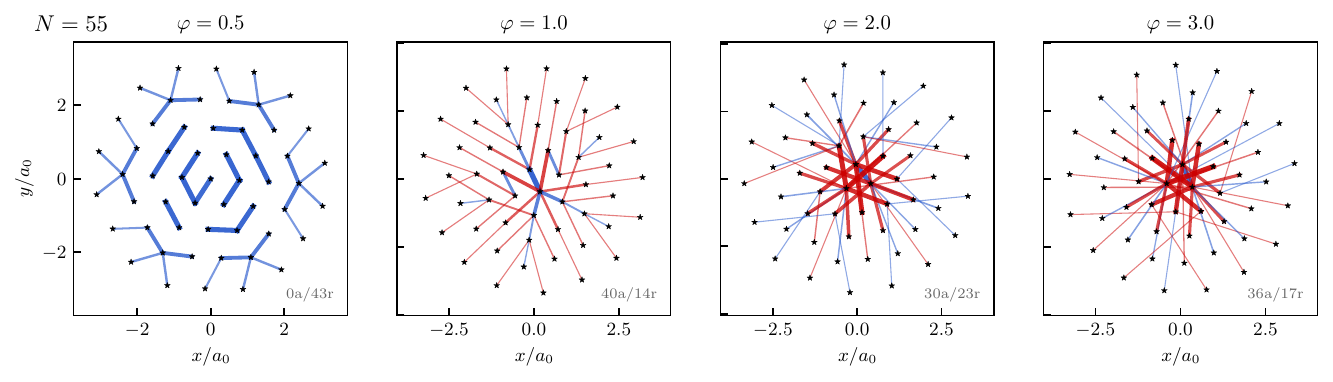}
\caption{Dominant pairwise force for $N=55$ at $\varphi=0.5$, $1.0$, $2.0$, $3.0$. Labels indicate attractive (a) / repulsive (r) bond counts. Attractive dominant bonds emerge and grow with decreasing temperature.}
\label{fig:SM_temp_N55}
\end{figure}

\begin{figure}[H]
\centering
\includegraphics[width=0.48\textwidth]{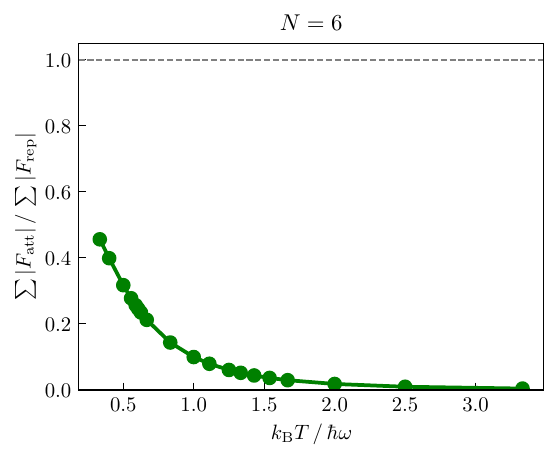}%
\hfill
\includegraphics[width=0.48\textwidth]{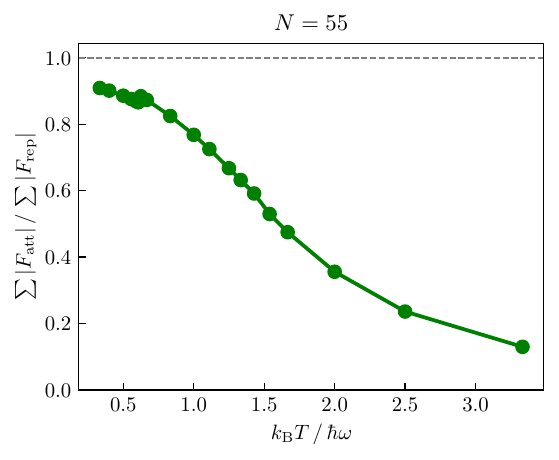}
\caption{Ratio $\sum|F_{\rm att}|/\sum|F_{\rm rep}|$ as a function of temperature for $N=6$ (left) and $N=55$ (right). The ratio grows monotonically with decreasing temperature, reflecting the strengthening of collective attraction relative to repulsion; for $N=55$ it approaches unity at low temperature. This ratio serves as a mechanical diagnostic of crystallization, tracking the structural order parameter $r_{\min}$ (Supplementary Note~5).}
\label{fig:SM_ratio}
\end{figure}

Supplementary Fig.~\ref{fig:SM_temp_N55} displays the dominant pairwise force for $N=55$ at the same four temperatures. At $\varphi=0.5$, all dominant forces are repulsive. As the temperature decreases, attractive dominant bonds emerge in the inner shells, with the ratio of attractive to repulsive bonds increasing monotonically (Supplementary Fig.~\ref{fig:SM_ratio}).

\begin{figure}[H]
\centering
\includegraphics[width=0.95\textwidth]{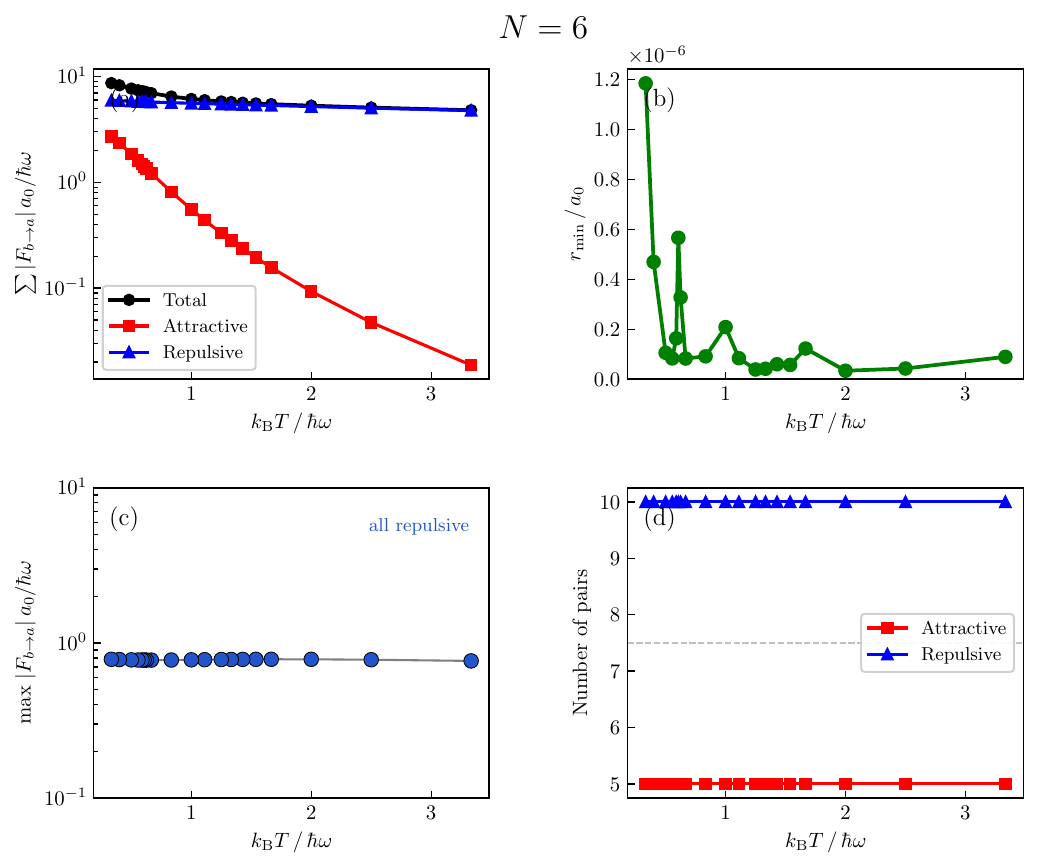}
\caption{Temperature dependence of pairwise statistical forces and structural order parameter for $N=6$ (companion to main text Fig.~4, which shows $N=55$); forces in units of $\hbar\omega/a_0$. (a)~Total, attractive, and repulsive force sums; (b)~$r_{\min}$; (c)~maximum pairwise force magnitude; (d)~number of attractive and repulsive pairs. The pair topology is fixed ($n_{\rm att}=5$, $n_{\rm rep}=10$) and the maximum force remains repulsive at all temperatures, in contrast to the crossover seen for $N=55$.}
\label{fig:SM_melting_N6}
\end{figure}

\begin{figure}[H]
\centering
\includegraphics[width=0.95\textwidth]{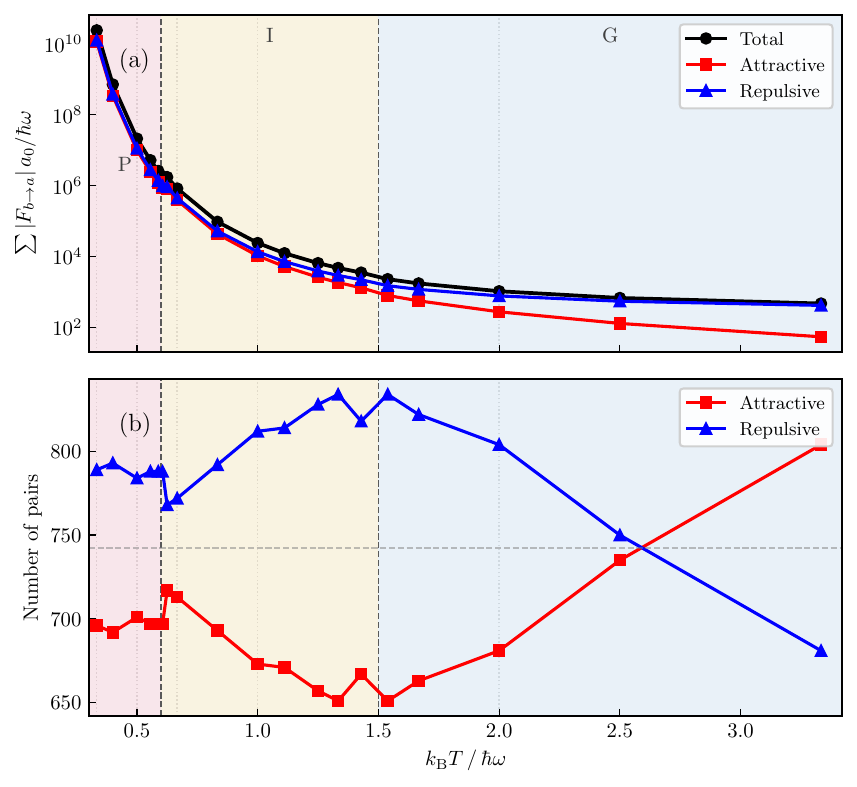}
\caption{Complementary force content for the $N=55$ temperature scan in main text Fig.~4, with the same P/I/G structural-regime shading. (a)~Total, attractive, and repulsive sums of pairwise-force magnitudes, in units of $\hbar\omega/a_0$. (b)~Numbers of attractive and repulsive pairs; the dashed horizontal line marks half of the $N(N-1)/2$ pairs. These global aggregates vary continuously within each structural regime but reorganize at the G$\to$I and I$\to$P level crossings.}
\label{fig:SM_melting_N55_aux}
\end{figure}

Supplementary Fig.~\ref{fig:SM_melting_N6} presents the full temperature dependence of the force content for $N=6$, the companion to main text Fig.~4: the pair topology is fixed ($n_{\rm att}=5$, $n_{\rm rep}=10$) and the maximum force remains repulsive at all temperatures, in contrast to the crossover of $N=55$. Supplementary Fig.~\ref{fig:SM_melting_N55_aux} gives the complementary $N=55$ force sums and pair counts omitted from the streamlined main figure. Supplementary Table~\ref{tab:Fmax_crossover} lists the strongest pairwise force, its type, and the structural order parameter for $N=55$ at representative temperatures.

\begin{table}[H]
\centering
\caption{Temperature dependence of the globally strongest pairwise force and structural order parameter for $N=55$.}
\label{tab:Fmax_crossover}
\vspace{4pt}
\begin{tabular}{ccccc}
\hline\hline
~~$k_{\rm B}T/\hbar\omega$~~ &~~ $|F|_{\rm max}\,a_0/\hbar\omega$~~ &~~ type~~ &~~ $r_{\min}/a_0$~~ &~~ phase~~ \\
\hline
0.33 & $3.9\times10^{8}$  & attractive & 0.42 & P \\
0.50 & $3.4\times10^{5}$  & attractive & 0.43 & P \\
0.67 & $2.2\times10^{4}$  & repulsive  & 0.31 & I \\
1.00 & $7.2\times10^{2}$  & repulsive  & 0.30 & I \\
2.00 & $1.0\times10^{1}$  & repulsive  & $\approx 0$ & G \\
\hline\hline
\end{tabular}
\end{table}

\FloatBarrier
\section*{Supplementary Note 5: Structural transitions\\ of the global minimum}
\label{SI:struct}

\begin{figure}[H]
\centering
\includegraphics[width=0.59\textwidth]{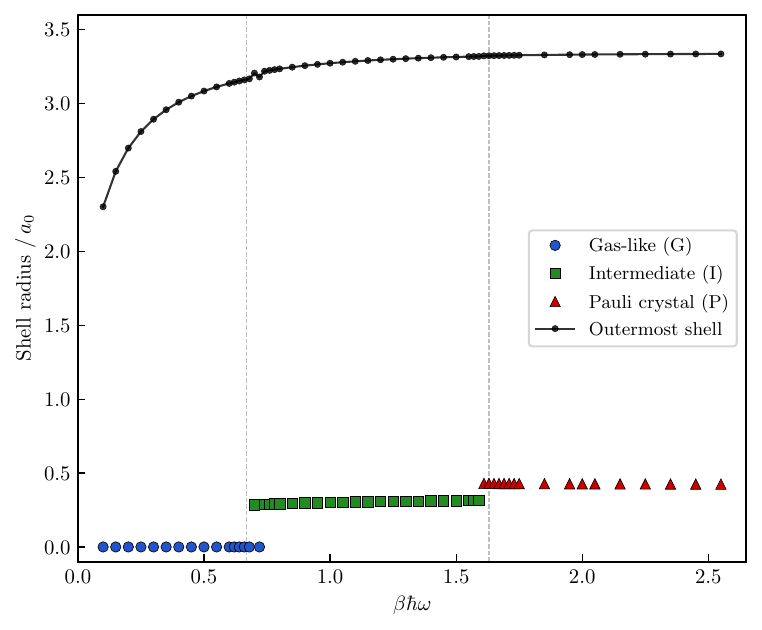}
\caption{Innermost-particle radius $r_{\min}(\varphi)=\min_a|\vx_a^\star|$ (colored markers) and outermost-shell mean radius (black) for $N=55$ as a function of $\varphi=\beta\hbar\omega$. Three structural phases are distinguished: gas-like~(G, blue), intermediate~(I, green), and Pauli crystal~(P, red). The discontinuous jumps of $r_{\min}$ at $\varphi\approx 0.67$ and $\varphi\approx 1.63$ indicate structural transitions, while the outermost-shell radius varies smoothly. The G and I phases are repulsion-dominated, whereas the P phase is attraction-dominated (cf.\ main text Fig.~4).}
\label{fig:SM_rmin}
\end{figure}

The global minimum of $V_{\rm total}(\bX)$ exhibits distinct structural phases as a function of temperature. For $N=55$, we define the order parameter
\be
r_{\min}(\varphi):=\min_{a}\,|\vx_{a}^{\star}(\varphi)|\,,
\label{rmindef}
\ee
the radius of the innermost particle at the $V_{\rm total}$ minimum (see also main text Fig.~4). Supplementary Fig.~\ref{fig:SM_rmin} shows $r_{\min}$ together with the outermost-shell mean radius as a function of $\varphi=\beta\hbar\omega$. Three regimes are clearly distinguished:

\begin{itemize}
\item \textbf{Gas-like~(G) phase} ($\varphi\lesssim 0.65$): a single particle sits at the origin, $r_{\min}\approx 0$.
\item \textbf{Intermediate~(I) phase} ($0.68\lesssim\varphi\lesssim 1.60$): the two innermost particles move away from the origin to $r_{\min}\approx 0.3\,a_0$, with the third-closest particle remaining at $r\approx 0.7\,a_0$.
\item \textbf{Pauli crystal~(P) phase} ($\varphi\gtrsim 1.65$): three particles form an equilateral triangle with $r_{\min}\approx 0.43\,a_0$, coinciding with the Pauli crystal geometry.
\end{itemize}

\begin{figure}[H]
\centering
\includegraphics[width=0.80\textwidth]{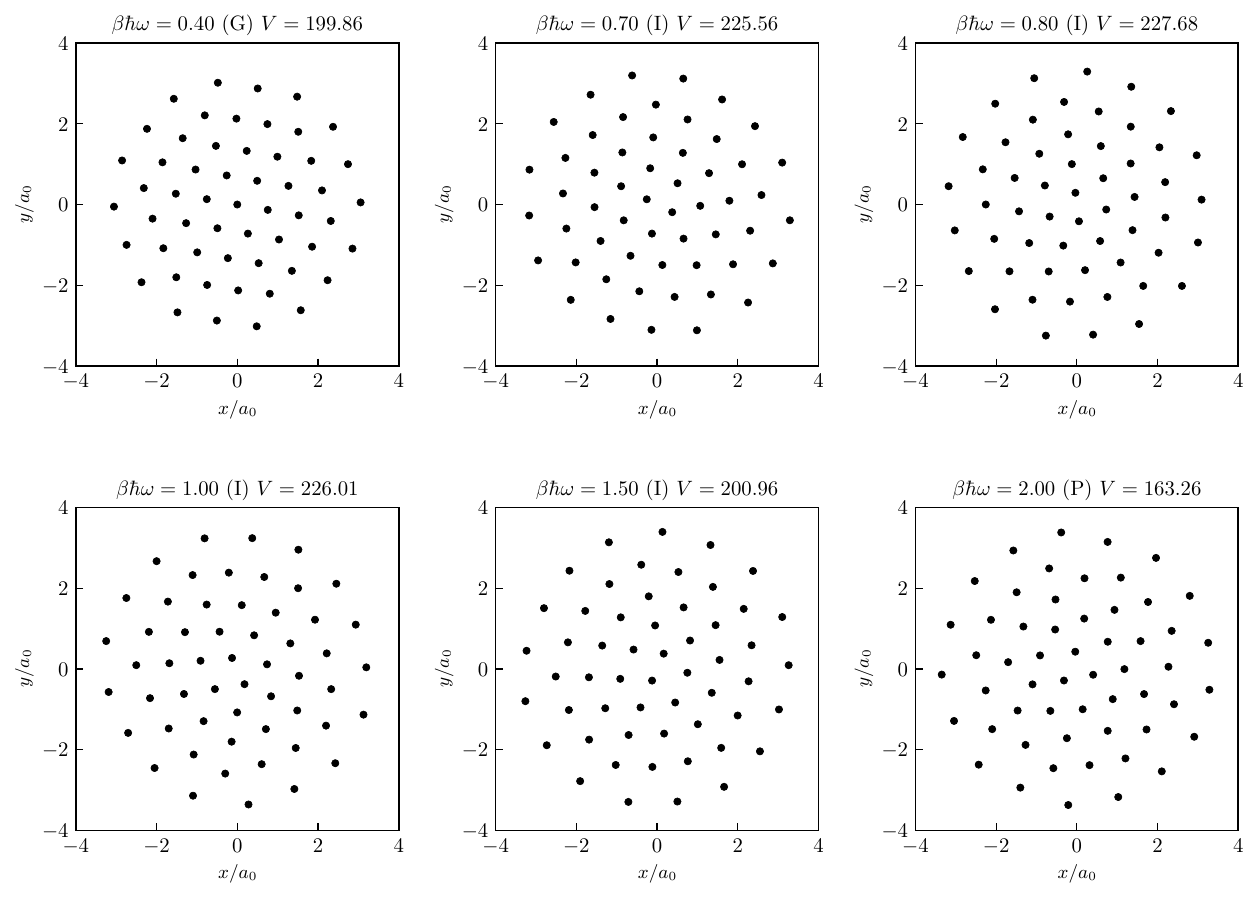}
\caption{Representative configurations at the $V_{\rm total}$ minimum for $N=55$. \textbf{Top row:} G~phase ($\varphi=0.40$) and I~phase ($\varphi=0.70$, $0.80$). \textbf{Bottom row:} I~phase ($\varphi=1.00$, $1.50$) and P~phase ($\varphi=2.00$). The innermost particle structure changes from one (G) to two (I) to three (P) particles.}
\label{fig:SM_struct_configs}
\end{figure}

The G$\to$I transition at $\varphi\approx 0.67$ ($k_{\rm B}T/\hbar\omega\approx 1.5$) and the I$\to$P transition at $\varphi\approx 1.63$ ($k_{\rm B}T/\hbar\omega\approx 0.62$) are both discontinuous in $r_{\min}$, while the outermost-shell radius varies smoothly (Supplementary Fig.~\ref{fig:SM_rmin}). Supplementary Fig.~\ref{fig:SM_struct_configs} shows representative configurations in each phase.

\begin{figure}[H]
\centering
\includegraphics[width=0.70\textwidth]{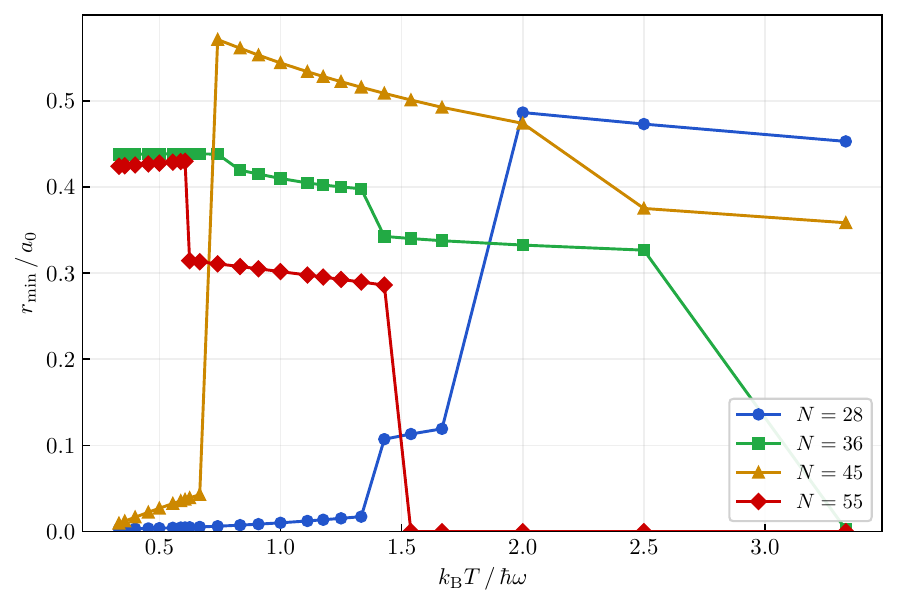}
\caption{Order parameter $r_{\min}=\min_a|\vx_a^\star|$ as a function of temperature for closed-shell particle numbers $N=28$, $36$, $45$, $55$. Each $N$ exhibits a distinct transition pattern: for $N=55$ the G$\to$I$\to$P sequence is clearly visible, $N=28$ and $N=45$ exhibit sharp transitions at different temperatures, and $N=36$ shows a milder crossover. Point-to-point alternation near the transition temperatures reflects competing minima with nearly degenerate energies in the nonconvex $V_{\rm total}$ landscape.}
\label{fig:SM_rmin_multiN}
\end{figure}

These are not phase transitions in the thermodynamic-limit sense: they are level crossings between competing local minima of the exact effective landscape. Near the transition temperatures the $V_{\rm total}$ landscape supports competing minima with nearly degenerate energies, which can cause the global minimum to alternate between structural phases over a narrow temperature range (Supplementary Fig.~\ref{fig:SM_rmin_multiN}). Direct Monte Carlo sampling of the thermal ensemble $e^{-\betavar V_{\rm total}}$ (Supplementary Note~8) shows, however, that this competition between minima does \textit{not} produce bimodality in the marginal distributions of the low-order radial order statistics: these marginals remain unimodal and nearly temperature-independent across both transitions, so the G$\to$I$\to$P sequence characterizes the distribution's mode rather than its bulk. The ensemble-level signature of the transitions is carried instead by the bond-sign statistics (Supplementary Note~8).

\subsection*{Constant central pressure}

Notably, these structural changes occur under constant central pressure. For any gas in a radially symmetric 2D harmonic trap at temperature $T$, the local force balance between the pressure gradient and the trap force on the number density $n(r,T)$ reads
\be
\partial_{r}P(r,T)=-m\omega^2 r\,n(r,T)\,.
\label{forcebalance}
\ee
Defining $N(r,T)={\int_{0}^{r}}n(r^{\prime},T)\,2\pi r^{\prime}\,{\rm{d}} r^{\prime}$ as the number of particles inside radius $r$ and integrating with the boundary condition $P(\infty,T)=0$,
\be
P(r,T)=\frac{m\omega^2}{2\pi}\big[N-N(r,T)\big]\,.
\label{Pr}
\ee
At the center, $N(0,T)=0$, so
\be
P(0,T)=\frac{m\omega^2}{2\pi}\,N\,,
\label{P0}
\ee
independently of temperature, quantum statistics, or interactions [main text Ref.~24]. The discontinuous rearrangement of the innermost particles at fixed central pressure is reminiscent of the isobar zigzags found in ideal Bose gases [main text Refs.~24, 25].

\subsection*{Stability of a central particle in the crystal phase}

The optimal occupancy of the trap center changes discretely along the G$\to$I$\to$P sequence: one particle at the origin in the G phase, none in the I phase (two innermost particles at $r_{\min}\approx0.3\,a_0$), and none in the P phase (three innermost particles forming a triangle at $r_{\min}\approx0.43\,a_0$). A configuration with a particle placed exactly at the trap center is always a stationary point of $V_{\rm total}$---the net statistical and trap forces on the central particle vanish by symmetry---but in the crystal regime it is not the global minimum: relaxing such a configuration lowers $V_{\rm total}$ by displacing the central particle and reorganizing the innermost shell into the triangular arrangement. In other words, central occupancy is favored only where the landscape makes it so (the G phase); in the P phase the centered configuration is a higher-lying (metastable or unstable) stationary point. This resolves the question of how the crystal phase accommodates a particle at the origin: it does not---any centrally placed particle is expelled to the triangular inner shell upon relaxation.

\FloatBarrier
\section*{Supplementary Note 6: Distance dependence of the statistical force}
\label{SI:dist}

\begin{figure}[H]
\centering
\includegraphics[width=0.65\textwidth]{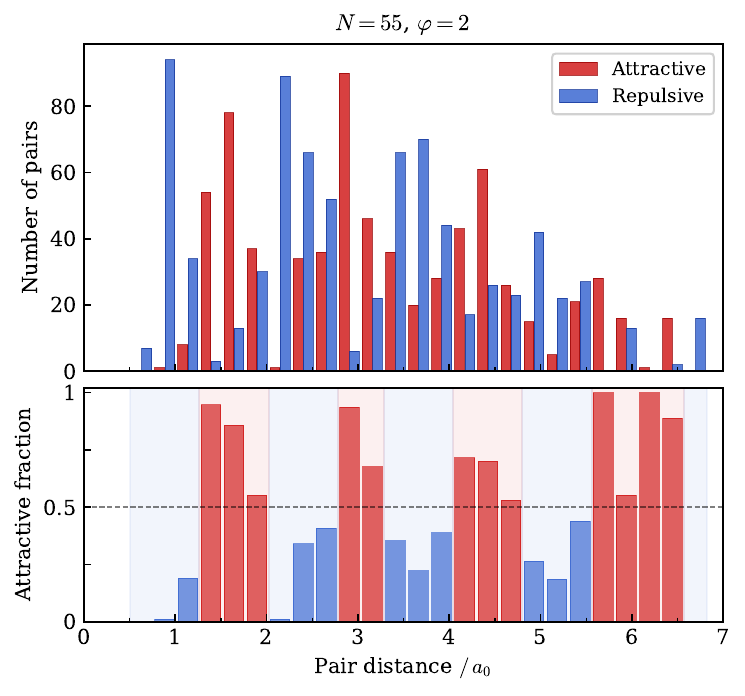}
\caption{Distance dependence of the statistical force type for $N=55$ at $\varphi=2$. \textbf{Top:} Number of attractive (red) and repulsive (blue) pairs in each distance bin. \textbf{Bottom:} Attractive fraction, showing oscillatory behavior across shells. The dashed line marks $50\%$.}
\label{fig:SM_dist_hist}
\end{figure}

Supplementary Fig.~\ref{fig:SM_dist_hist} shows the distribution of attractive and repulsive pair forces as a function of interparticle distance for $N=55$ at $\varphi=2$. The attractive fraction oscillates with distance, alternating between attraction-dominated and repulsion-dominated regimes. The nearest-neighbor pairs ($r/a_0\lesssim 1$) are predominantly repulsive, while the next shell of neighbors ($1\lesssim r/a_0\lesssim 2$) is predominantly attractive. This alternating pattern continues to larger distances, providing direct evidence for the shell-to-shell organization of the dual force-balance architecture described in the main text.

\FloatBarrier
\section*{Supplementary Note 7: Canonicality of the pairwise decomposition}
\label{SI:canonical}

Here we make precise in what sense the pairwise decomposition of the statistical force [main text Eqs.~(3)--(5)] is canonical, and why its uniqueness must be stated at the level of the representation.

\subsection*{The canonical decomposition}

The statistical potential is defined through the determinant representation
\be
V_{\rm stat}=-\beta^{-1}\ln\Det K\,,\qquad
K_{ab}=\exp\!\Big[-\tfrac{m}{2\beta\hbar^{2}}\,s_{ab}\Big]\,,\qquad
s_{ab}:=(\vx_a-\vx_b)^2\,,
\label{SIrep}
\ee
in which the configuration enters \textit{only} through the squared separations $s_{ab}$, each appearing in the single symmetric entry pair $K_{ab}=K_{ba}$. Treating the $s_{ab}$ ($a<b$) as independent arguments of this representation, the chain rule gives
\be
\vF_{a}=-\na_{a}V_{\rm stat}
=\sum_{b\neq a}\,2\,(\vx_{b}-\vx_{a})\,\frac{\partial V_{\rm stat}}{\partial s_{ab}}
=:\sum_{b\neq a}\vF_{b\rightarrow a}\,,
\qquad
\vF_{b\rightarrow a}=\frac{2m}{(\beta\hbar)^{2}}\,(\vx_{b}-\vx_{a})\,K^{-1\,ab}K_{ab}\,,
\label{SIchain}
\ee
using $\partial\ln\Det K/\partial s_{ab}=2\,K^{-1\,ab}\,\partial K_{ab}/\partial s_{ab}$ (the factor 2 from the symmetric entry pair; no sum over $a,b$). Each term is central (directed along $\vx_b-\vx_a$) and antisymmetric under $a\leftrightarrow b$ (Newton's third law), since $K$ and $K^{-1}$ are symmetric. This is the decomposition used throughout the main text. It is permutation-covariant, real-analytic in the $s_{ab}$, independent of any configuration-specific data, and reduces to the Uhlenbeck--Gropper force at $N=2$.

\subsection*{Uniqueness: exact for $N=3$, representation-level for $N\geq4$}

Consider two central, Newton's-third-law decompositions of the same total force field, $\vF_a=\sum_{b\neq a}c_{ab}(\bX)(\vx_b-\vx_a)$ with symmetric coefficients $c_{ab}=c_{ba}$. Their difference $\delta c_{ab}$ satisfies
\be
\sum_{b\neq a}\delta c_{ab}\,(\vx_{b}-\vx_{a})=0\quad\text{for all }a\,,
\label{selfstress}
\ee
which is precisely the defining equation of a \textit{self-stress} (equilibrium stress) of the complete graph on the points $\{\vx_a\}$ in the sense of rigidity theory. For $N$ generic points in $D=2$ dimensions, nonzero self-stresses exist if and only if the number of pair constraints exceeds the framework's degrees of freedom, $N(N-1)/2>2N-3$, i.e.\ for $N\geq4$. Equivalently: for $N\geq4$ the squared separations are functionally dependent (Cayley--Menger constraints), so a potential expressed as a function of the $s_{ab}$ admits different off-shell extensions whose partial derivatives differ exactly by self-stresses.

Consequently:
\begin{itemize}
\item For $N=3$ (generic, non-collinear configurations), the central pairwise decomposition is \textit{strictly unique}. In particular, the antipodal-ball criterion for the sign change of $\vF_{2\to1}$ [main text Eq.~(9)] is representation-independent.
\item For $N\geq4$, uniqueness holds at the level of the representation: among all central decompositions, Eq.~(\ref{SIchain}) is the unique one whose coefficients are the partial derivatives of the determinant representation~(\ref{SIrep}) with respect to the squared separations. This representation is the natural analytic extension---it is the closed form in which $V_{\rm stat}$ arises from the trace formula in the first place---and it is the canonical choice adopted throughout. All statements in the main text (sign classification of bonds, dominant-bond identification, the force crossover) refer to this canonical decomposition.
\end{itemize}

This clarification does not affect any quantity derived from the total force or from $V_{\rm stat}$ itself (minima, structural transitions, thermodynamics), which are manifestly decomposition-independent.

\subsection*{Numerical verification}

The statements above have been verified numerically (scripts are provided with the code repository). (i)~For random planar configurations the dimension of the self-stress space matches $(N-2)(N-3)/2$ exactly for $N=3,\ldots,8$; in particular it vanishes for $N=3$. (ii)~For $N=4$, adding the explicit self-stress $t\,\mu_a\mu_b$ built from the affine dependence $\sum_a\mu_a=0$, $\sum_a\mu_a\vx_a=0$ flips the sign of a chosen pair coefficient while changing no total force $\vF_a$ beyond the $10^{-13}$ level (machine precision). (iii)~At the $V_{\rm total}$ minima studied in the main text ($\varphi=2$, closed shells up to $N=55$, with the Table~1 shell radii reproduced to $2\times10^{-4}\,a_0$), linear programming over the self-stress space shows that \textit{pointwise} re-decompositions of the same total statistical force field exist in which every pair is repulsive---for the $N=55$ crystal with all coefficients bounded away from zero (uniform margin $\approx 7.6\times10^{-3}\,\hbar\omega/a_0^{2}$, equality residual at the $10^{-12}$ level after least-squares refinement). For $N\geq4$, even the existence of attractive components is therefore a representation-dependent statement at any single configuration; all sign classifications in this work refer to the canonical decomposition~(\ref{SIchain}), and the $N=3$ criterion of the main text is unconditional.

\subsection*{Scheme-level rigidity from the three-body sector}

The pointwise freedom above does not, however, extend to a decomposition \textit{scheme}. Consider any assignment of central, Newton's-third-law pair forces defined for all configurations, depending continuously on the configuration, and decaying for infinitely separated pairs. When $N-3$ particles are removed to infinity, the total statistical force field on the remaining triple converges to the $N=3$ field (a property of $V_{\rm stat}$ itself), while force decay eliminates the far-pair contributions; the induced decomposition on the triple must then converge to the \textit{unique} $N=3$ central decomposition---which is attractive inside the antipodal ball [main text Eq.~(9)]. No such scheme can therefore be everywhere repulsive: attraction within the statistical force is unavoidable for any continuous, cluster-separable pairwise representation. The all-repulsive constructions of the previous paragraph evade this only by being configuration-local. This sharpens the sense in which the attraction reported in the main text is a genuine feature of the many-body statistical force rather than an artifact of the canonical choice.

\section*{Supplementary Note 8: Thermal-ensemble Monte Carlo}
\label{SI:mc}

The strictly positive weight of the dual (main text) permits direct Monte Carlo sampling of the exact configurational distribution $e^{-\betavar V_{\rm total}}$. The sampler, its validation against the exact $N=2$ solution, and its convergence diagnostics are described in Methods. Here we report the ensemble-level results for $N=55$---three independent replica-exchange runs initialized from gas-like, crystalline, and mixed configurations, with split-chain $\hat R\le 1.001$ and effective sample sizes $\gtrsim 8{,}000$ per temperature---together with the $N=6$ null control.

\begin{figure}[H]
\centering
\includegraphics[width=0.95\textwidth]{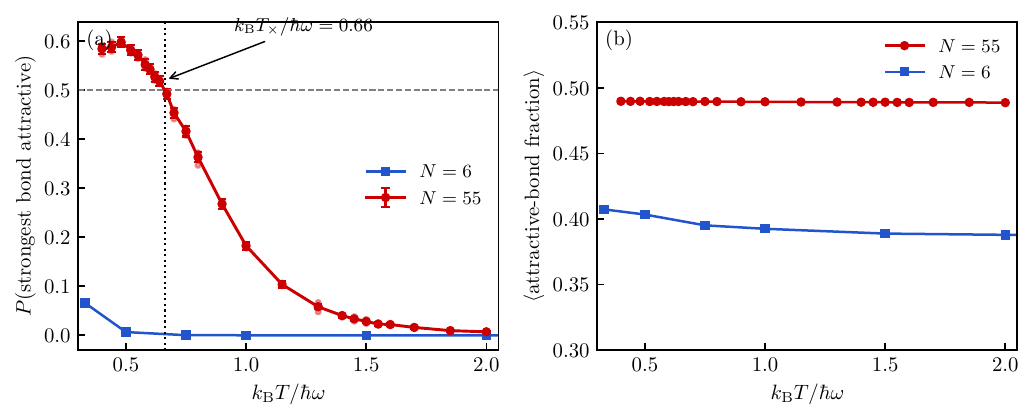}
\caption{Ensemble Monte Carlo of the exact dual. (a)~Probability that the strongest canonical bond of a single sampled configuration is attractive, for $N=55$ (red: three independent runs as light points; pooled values with Wilson $95\%$ intervals) and $N=6$ (blue). The $N=55$ curve crosses one half at $k_{\rm B}T_\times/\hbar\omega=0.66(1)$ (dotted); the $N=6$ null control never exceeds $0.07$. (b)~Mean attractive-bond fraction, nearly temperature-independent for both $N$.}
\label{fig:SM_mc_crossover}
\end{figure}

\textbf{Ensemble force crossover.} Supplementary Fig.~\ref{fig:SM_mc_crossover}(a) shows the probability that the strongest canonical bond of a single sampled configuration is attractive. For $N=55$ it rises smoothly from $7\times10^{-3}$ at $k_{\rm B}T/\hbar\omega=2.0$ to $0.59$ at $0.40$, crossing one half at $k_{\rm B}T_\times/\hbar\omega=0.66$ with run-to-run spread $0.003$ (per-run crossings $0.660$, $0.660$, $0.666$). This is the ensemble counterpart of the mode-level crossover at ${\approx}0.6$ (main text Fig.~4): a majority statistic of single shots, not merely a property of the optimal configuration. For $N=6$ the probability never exceeds $0.07$ at any temperature---the null control holds at ensemble level. The mean attractive-bond fraction [Supplementary Fig.~\ref{fig:SM_mc_crossover}(b)] is nearly temperature-independent ($\approx0.49$ for $N=55$, $\approx0.39$ for $N=6$): the crossover is carried by the identity and strength of the strongest bond, not by wholesale sign conversion of the bond network.

\begin{figure}[H]
\centering
\includegraphics[width=0.95\textwidth]{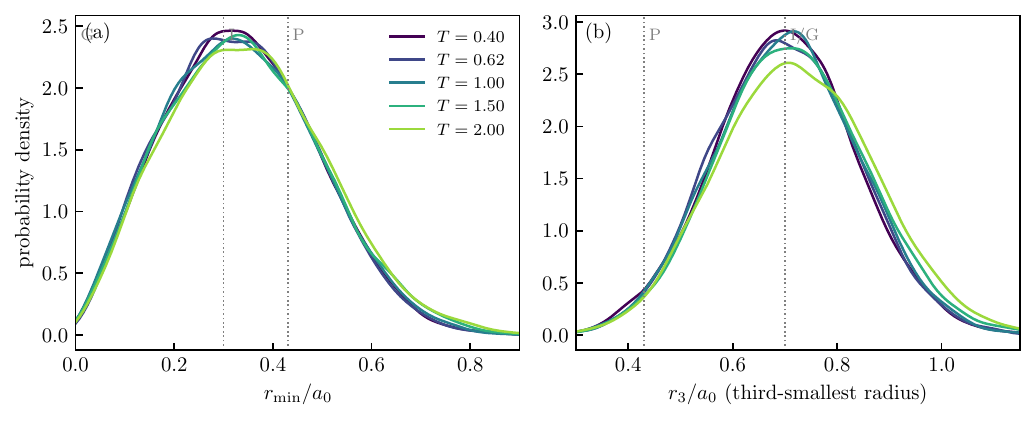}
\caption{Marginal distributions of the first and third radial order statistics across the structural transitions ($N=55$, pooled over three independent runs). Kernel-density estimates of (a)~the smallest radius $r_{\min}$ and (b)~the third-smallest radius $r_3$ at five temperatures spanning both transitions. Dotted lines mark the corresponding values at the global minimum in the G, I, and P regimes. The marginals are unimodal at every temperature---stable under bandwidth variation and reproducible across runs---and nearly temperature-independent: the discrete G$\to$I$\to$P rearrangements of the minimum are not accompanied by bimodality of these ensemble marginals.}
\label{fig:SM_mc_marginals}
\end{figure}

\textbf{Mode-level character of the structural transitions.} Supplementary Fig.~\ref{fig:SM_mc_marginals} shows kernel-density estimates of the marginal distributions of the first and third radial order statistics ($r_{\min}$ and $r_3$) across both structural transitions. At every temperature these marginals are unimodal---stable under kernel-bandwidth variation ($0.8\times$--$1.4\times$ Silverman) and reproducible across the three independent runs---and nearly temperature-independent, in sharp contrast with the discontinuous jumps of the same quantities evaluated at the global minimum (Supplementary Note~5). The G$\to$I$\to$P sequence is therefore a property of the distribution's mode: thermal broadening, comparable to the spacing between the competing minima, together with the exactly constant central pressure (Supplementary Note~5), leaves the bulk of the ensemble unimodal. Predictions for single-shot experiments should accordingly be phrased at the level of bond-sign statistics [Supplementary Fig.~\ref{fig:SM_mc_crossover}], which carry a strong, monotonic temperature dependence, rather than at the level of radial histograms.

\end{document}